\documentclass[a4paper,12pt]{article}  

\usepackage{graphicx}
\usepackage{amssymb}
\usepackage{amsmath} 
\usepackage{tensor}

\title{Divergent part of the stress-energy tensor for Maxwell's theory in curved 
space-time: a systematic derivation}

\date{\today}

\author{Roberto Niardi ORCID: 0000-0001-9216-3322, \\
	Dipartimento di Fisica \lq\lq Ettore Pancini'', \\
	Universit\`a degli Studi di Napoli Federico II, Italy
	\and 
	Giampiero Esposito ORCID: 0000-0001-5930-8366 \\
        Dipartimento di Fisica \lq\lq Ettore Pancini'', \\
        Universit\`a degli Studi di Napoli Federico II, Italy \\
        Complesso Universitario di Monte S. Angelo, \\ 
	Via Cintia Edificio 6, 80126 Napoli, Italy, \\
        Istituto Nazionale di Fisica Nucleare, Sezione di
	Napoli, \\ 
	Complesso Universitario di Monte S. Angelo, \\ 
	Via Cintia Edificio 6, 80126 Napoli, Italy
	\and
	Francesco Tramontano ORCID: 0000-0002-3629-7964
	\\
    Dipartimento di Fisica \lq\lq Ettore Pancini'', \\
    Universit\`a degli Studi di Napoli Federico II, Italy \\
    Complesso Universitario di Monte S. Angelo, \\ 
    Via Cintia Edificio 6, 80126 Napoli, Italy, \\
    Istituto Nazionale di Fisica Nucleare, Sezione di
    Napoli, \\ 
    Complesso Universitario di Monte S. Angelo, \\ 
    Via Cintia Edificio 6, 80126 Napoli, Italy
}

\begin{document}
\maketitle

\begin{abstract}
In this paper the Feynman Green function for Maxwell's theory in curved space-time is studied 
by using the Fock-Schwinger-DeWitt asymptotic expansion; the point-splitting method is then 
applied, since it is a valuable tool for regularizing divergent observables. Among these, 
the stress-energy tensor is expressed in terms of second covariant derivatives 
of the Hadamard Green function, which is also closely linked to the effective action; therefore 
one obtains a series expansion for the stress-energy tensor. Its divergent part can be isolated, 
and a concise formula is here obtained: by dimensional analysis and combinatorics, there are two kinds 
of terms: quadratic in curvature tensors (Riemann, Ricci tensors and scalar curvature) and linear 
in their second covariant derivatives. This formula holds for every space-time metric; it is made 
even more explicit in the physically relevant particular cases of Ricci-flat and maximally symmetric 
spaces, and fully evaluated for some examples of physical interest: Kerr and Schwarzschild metrics  
and de Sitter space-time.
\end{abstract}

\section{Introduction}
\setcounter{equation}{0}

The study of quantum gauge field theories and gravitation is a necessity connected to black hole 
theory and quantum cosmological models, but beyond that it is also a crucial intellectual achievement.
On one hand, electroweak and strong interactions are described by Yang-Mills theories, and are 
satisfactorily set in a quantum framework; on the other hand, altough General Relativity is in many 
ways a gauge theory too \cite{dewitt2003global}, it stands apart from the other three forces of 
nature, and \textit{a quantum theory of gravity does not yet exist}, at least as a coherent discipline.
Nevertheless, it is possible to discuss the influence of the gravitational field on quantum phenomena: 
one can study the regime for quantum aspects of gravity in which the gravitational field 
is described as a classical background through Einstein's theory while 
matter fields are quantized; this is reasonable as long as length and time scales of 
quantum processes of interest are greater than the Planck values ($l_{Planck} \equiv (G\hbar /c^3)^{1/2} 
\sim 1.616 \times 10^{-33}{\rm cm}$, $t_{Planck} \equiv (G\hbar /c^5)^{1/2} \sim 5.39 \times 10^{-44}{\rm s}$). 
Since Planck length is so small (twenty orders of magnitude below the size of an atomic nucleus), 
one can hope that such a \lq\lq semiclassical" approach has some predictive power. Therefore one 
is naturally led to the subject of quantum field theory in a curved background spacetime. Its 
basic physical prediction is that \textit{strong gravitational fields can \lq\lq polarize" the 
vacuum and, when time-dependent, lead to pair creation}; moreover, in a curved space-time, notions 
of \lq\lq vacuum" and \lq\lq particles" need a deeper discussion than in the flat case. These two 
fundamental results are strongly linked to the most important predictions of the theory, i.e., 
Hawking and Unruh effects (see Refs. \cite{hawking1974black}, \cite{hawking1975particle}, 
\cite{unruh1976notes}): according to the Hawking effect, a classical, spherically symmetric 
black hole of mass $M$ has the same emission spectrum of a black body having the 
temperature $T_{Hawking} \equiv \frac{1}{8\pi M}$; according to the Unruh effect, from 
the point of view of an accelerating observer, empty space contains a gas of particles 
at a temperature proportional to the acceleration. For a detailed treatise on these subjects, see 
also Refs. \cite{birrell1984quantum}, \cite{fulling1989aspects}, \cite{parker2009quantum}.

In this context, DeWitt's formalism for gauge field theories is a powerful tool (see Refs.  
\cite{dewitt1965dynamical}, \cite{dewitt1975quantum1}, \cite{de1983houches}, 
\cite{dewitt2003global}): it provides a 
framework in which the quantization of fields possessing infinite dimensional invariance groups may 
be carried out in a manifestly covariant fashion, even in curved space-time; moreover, it makes it 
possible to classify all gauge theories in a purely geometrical way, i.e., by the 
algebra which the generators of the gauge group obey; the geometry of such theories is the 
fundamental reason underlying the emergence of ghost fields in the corresponding quantum theories, too. 

For a non-gauge theory, it is well known that the transition amplitude $\langle {\rm out }|{\rm in } 
\rangle$ between any two (super)state vectors determined by sone unspecified boundary 
conditions may be expressed as 
\begin{eqnarray}
& &\langle {\rm out }|{\rm in } \rangle = N \int  e^{iS[\phi]}\mu[\phi] [d\phi], \\
& &[d\phi] \equiv \prod_{i} d\phi^i . 
\label{field_measure}
\end{eqnarray}
where $N$ is a normalization factor, while $\mu[\phi]$ is a suitable 
measure functional.\footnote{Altough this formula, for a general 
non-gauge theory, can be derived in a highly formal manner, rigorous results concerning functional 
integration and path integral can be found in the textbooks \cite{simon1979functional}, 
\cite{glimm2012quantum}, \cite{cartier2006functional}.}
DeWitt regards the set of all fields as an infinite-dimensional manifold, the 
space of histories, on which an action functional is defined; then gauge transformations are viewed 
as flows which leave the action functional unchanged; when the flows $Q_\alpha$ satisfy the equations 
of a closed (altough infinite-dimensional) Lie group (this is the case for Yang-Mills theories and 
General Relativity), under the group action the space of histories is separated into orbits; 
one could say that it is in the space of orbits that the real physics of the system takes place; 
one can then choose on the manifold a new set of coordinates made of two parts: the first 
one labelling the orbit, $I^A$, and the second one, $K^\alpha$, labelling a particular field 
configuration belonging to the specified orbit; when one writes down the functional integral and 
reverts to the original coordinates, one finds out that two new terms appear: 
\begin{eqnarray}
&&\langle {\rm out }|{\rm in } \rangle = \int \mu[\phi] [d\phi] \int [d\chi] \int [d\psi] \; 
e^{i(S[\phi]+\Omega + \chi_\alpha \mathcal{M}^\alpha _\beta \psi^\beta)}, \label{path_int_ghost} \\
&& \Omega \equiv \frac{1}{2} \kappa_{\alpha\beta}K^\alpha K^\beta, \\
&& \mathcal{M}^\alpha _\beta \equiv K^\alpha Q_\beta ,
\end{eqnarray}
where $\Omega \equiv \frac{1}{2} \kappa_{\alpha\beta}K^\alpha K^\beta$ is the 
gauge-breaking (or gauge-averaging) term, while $\chi_\alpha \mathcal{M}^\alpha _\beta \psi^\beta$ 
is the ghost contribution, which involves two ghost fields; they are fermionic for a 
bosonic theory and bosonic for a fermionic theory.
Therefore one refers to 
\begin{equation}
\bar{S} \equiv S[\phi]+ \frac{1}{2} \kappa_{\alpha\beta}K^\alpha K^\beta 
+ \chi_\alpha \mathcal{M}^\alpha _\beta \psi^\beta 
\end{equation}
as the full action functional.

In this paper the Feynman Green function for Maxwell's theory in curved space-time is discussed, 
following the footsteps of DeWitt and \cite{christensen1978regularization} and \cite{bimonte2003photon}: 
both $\zeta$-function regularization method and Fock-Schwinger-DeWitt ansatz are used in order to 
obtain an asymptotic formula for the Feynman propagator, valid for small values of the geodetic 
distance; the expansion obtained exhibits the familiar logarithmic singularity which occurs for 
massive theories in flat space-time and, generally, even for massless theories but in curved 
space-time. Then the point-splitting method is presented as a valuable tool for regularizing 
divergent observables such as the stress-energy tensor: one finds out that it can be completely 
expressed in terms of second derivatives of the Hadamard Green function, which is also closely 
linked to the effective action. Most of the material in Secs. 2, 3 is quite standard, but 
nevertheless necessary to obtain a self-contained presentation.
In Sec. 4 an original calculation is presented: a short, closed formula for the divergent part 
of the stress-energy tensor, originating from a careful handling of more than two thousand terms; 
for this purpose, a program has been written in FORM, which is a symbolic manipulation system 
whose original author is Jos Vermaseren at NIKHEF (see Refs. 
\cite{Heck1}, \cite{Vermaseren1}, \cite{courses1}). 

This formula holds for every space-time metric; in Secs. 5 and 6, it is made even more explicit 
in the physically relevant particular cases of Ricci-flat and maximally symmetric spaces, and fully 
evaluated for some relevant examples: Kerr and Schwarzschild metrics (which are Ricci-flat) and 
de Sitter metric (which is maximally symmetric); see Refs. \cite{teukolsky2015kerrmetric}, 
\cite{realdi2008einstein} for an up-todate review. Concluding remarks are presented in Sec. 7,
and relevant details are given in the Appendices.

\section{Quantum Maxwell Theory in Curved Space-Time and Feynman Propagator}
\setcounter{equation}{0}

In this section, the task will be the evaluation of the photon Green's functions in curved space-time. 
Following DeWitt's approach and using the minimal coupling, the full action functional for the theory is
(our spacetime metric has signature $(-,+,+,+)$)
\begin{equation}
\bar{S} = \int d^4 x \; | \mathrm{ g } |^{1/2} \left(-\frac{1}{4}F_{\mu\nu}F^{\mu\nu}
-\frac{\left(\tensor{A}{^\mu_{;\mu}}\right)^2}{2\xi}-\frac{\chi \Box \psi}{\sqrt{\xi}}\right). 
\label{full_ed_act}
\end{equation}
The first term is the classical action functional for the field $A_\mu$, while the second and the 
third ones are the gauge-averaging and the ghost ones, respectively; moreover, $\xi$ is a real 
parameter which accounts for the behaviour under rescaling of the gauge-fixing coordinate $K^\alpha$:
\begin{eqnarray}
&&K^\alpha \mapsto \frac{1}{\sqrt{\xi}}  K^\alpha 
\nonumber \\
&\;& \Longrightarrow \;
\begin{cases} \Omega \mapsto  \frac{1}{\xi}\; \Omega, \\
M^{\alpha}_\beta \mapsto \frac{1}{\sqrt{\xi}} M^{\alpha}_\beta.
\end{cases}
\end{eqnarray}
The full action functional \eqref{full_ed_act} can be put in the form
\begin{equation}
\bar{S} = \int d^4 x \; | \mathrm{ g } |^{1/2} \left( -\tfrac{1}{2} A_\mu P^{\mu\nu}(\xi) A_\nu 
+ \tfrac{1}{\sqrt{\xi}}\chi P_0 \psi \right),
\end{equation}
where 
\begin{eqnarray}
P^{\mu\nu}(\xi) &=& -g^{\mu\nu} \Box +R^{\mu\nu} + (1-\tfrac{1}{\xi})\nabla^\mu \nabla^\nu, \\
P_0 &=& -\Box = -g^{\nu\mu}\nabla_\mu \nabla_\nu.
\end{eqnarray}
Hence the equation for the photon Green's functions is
\begin{equation}
| \mathrm{ g } |^{1/2} P_\mu^\lambda(\xi) G^{(\xi)}_{\lambda\nu}(x,x') = g_{\mu\nu}(x) \delta(x,x').
\end{equation}
In order to study these equations, we introduce two non-physical Hilbert spaces spanned by the basis 
kets $|x,\mu\rangle$, $|x \rangle$, respectively; for them, the following orthogonality relations hold:
\begin{eqnarray}
\langle x,\mu | x',\nu \rangle &=& g_{\mu\nu}(x) \delta(x,x'),  \\
\langle x| x'\rangle &=&  \delta(x,x').
\end{eqnarray}
Of course, $|x \rangle$ is the familiar Dirac notation for the eigenfunctionals of the position 
operator which has continuous spectrum, while the index of both $|x,\mu\rangle$ and the associated 
\lq\lq bra'' $\langle x,\mu |$ is viewed as that of a covariant vector density of weight $1/2$. 
The \lq\lq Hamiltonian'' operators $H(\xi)$, $H_0$ associated with $P^{\mu\nu}(\xi)$ 
and $P_0$, respectively, are defined by 
\begin{eqnarray}
\langle x,\mu |H(\xi)| x',\nu \rangle &=& P^\lambda_\mu (\xi) \langle x,\lambda| x',\nu \rangle, \\
\langle x |H_0| x' \rangle &=& P_0\langle x | x' \rangle.
\end{eqnarray}
Then the \lq\lq operator'' solution with the Feynman prescription is 
\begin{equation}
|\mathrm{g}|^{1/4} G^{(\xi)} |\mathrm{g}|^{1/4} = \frac{1}{H(\xi)-i\epsilon} 
= i\int d\tau \; e^{-i\tau H(\xi)}. 
\label{op_solution_F_prescr}
\end{equation}
By taking matrix elements of the previous equation we obtain 
\begin{eqnarray}
&&|\mathrm{g(x)}|^{1/4} G^{(\xi)}_{\mu\nu}(x,x') |\mathrm{g}(x')|^{1/4} = i \int _0 ^{+\infty} d\tau \; 
\left\langle x,\mu |e^{-i\tau H(\xi)}|x',\nu \right\rangle^{(\xi)} 
\nonumber \\
&&\;\;\;\;\;=  i \int _0 ^{+\infty} d\tau \; \left\langle x,\mu;\tau |x',\nu;0 \right\rangle^{(\xi)}, 
\label{massless_limit}
\end{eqnarray}
where 
\begin{equation}
\left\langle x,\mu;\tau |x',\nu;0 \right\rangle \equiv \left\langle x,\mu 
|e^{-i\tau H(\xi)}|x',\nu \right\rangle.
\end{equation}
Thus the \lq\lq transition amplitude'' $\left\langle x,\mu;\tau |x',\nu;0\right\rangle ^{(\xi)}$ 
satisfies the Schr\"{o}dinger equ\-at\-io\-n associated to $H(\xi)$:
\begin{equation}
i\frac{\partial}{\partial \tau} \left\langle x,\mu;\tau |x',\nu;0\right\rangle ^{(\xi)} 
=  \left\langle x,\mu;\tau |H(\xi) |x',\nu;0\right\rangle ^{(\xi)}, 
\label{tr_ampl_eq}
\end{equation}
with the initial condition 
\begin{equation}
\left\langle x,\mu;0 |x',\nu;0\right\rangle ^{(\xi)}= g_{\mu\nu}(x) \delta(x,x').
\end{equation}
For arbitrary values of $\xi$, the operator $P^{\mu\nu}(\xi)$, as well as $P^\lambda _\mu \equiv 
g_{\rho\mu}P^{\rho\lambda}(\xi)$, is non-minimal, i.e. the wavelike-operator part $-g^{\mu\nu}\Box 
+ R^{\mu\nu}$ is spoiled by $(1-1/\xi)\nabla^\mu \nabla^\nu$. Nevertheless, if one knows 
$\left\langle x,\mu;\tau |x',\nu;0\right\rangle ^{(\xi)}$ at $\xi = 1$, one can use this solution, 
here denoted by $\left\langle x,\mu;\tau |x',\nu;0\right\rangle ^{(1)}$, to evaluate 
$\left\langle x,\mu;\tau |x',\nu;0\right\rangle ^{(\xi)}$ according to the Endo formula 
(see Ref. \cite{endo1984gauge})
\begin{equation}
\left\langle x,\mu;\tau |x',\nu;0\right\rangle ^{(\xi)} = \left\langle x,\mu;\tau 
|x',\nu;0\right\rangle ^{(1)} + i \int_{\tau} ^{\tau/\xi} dy \; \nabla_\mu \nabla^\lambda 
\left\langle x,\lambda;y |x',\nu;0\right\rangle ^{(1)}. 
\label{Endo}
\end{equation}
The previous equation plays a key role in evaluating the regularized photon Green function, 
as we shall see in the following.

Once the transition amplitude is known, equation \eqref{massless_limit} describes the 
massless limit of the Feynman propagator (for which one would have to add an infinitesimal 
negative imaginary mass). At this stage, formula \eqref{massless_limit} needs a suitable 
regularization. Following Endo, we use $\zeta$-function regularization and introduce a 
regularization parameter $\mu_{A}$ defining (any suffix to denote regularization of the photon 
Green function is omitted for simplicity of notation):
\begin{equation}
|\mathrm{g(x)}|^{1/4} G^{(\xi)}_{\mu\nu}(x,x') |\mathrm{g}(x')|^{1/4} \equiv \lim_{s \to 0} 
\frac{\mu_A ^{2s} \; i^{s+1}}{\Gamma(s+1)} \int_0 ^{+\infty} d\tau \; \tau^s \left\langle x,\mu;
\tau |x',\nu;0 \right\rangle^{(\xi)}. 
\label{reg_ph_Gr}
\end{equation}
It should be stressed that the limit $s \to 0$ should be taken at the very end of all 
calculations, and cannot be brought within the integral.

The transition amplitude $\left\langle x,\mu;\tau |x',\nu;0\right\rangle ^{(1)}$ is known as 
$\tau \to 0$ and as $$\sigma(x,x') \to 0$$ (where the \textit{bi-scalar} $\sigma (x,x')$, 
which we shall call \textit{geodetic interval} or \textit{world function}, is equal to  
half the square of the distance along the geodesic between $x$ and $x'$) through its 
Fock-Schwinger-DeWitt asymptotic expansion (Refs. \cite{dewitt1965dynamical}, \cite{dewitt2003global}, 
\cite{fock1937eigen}, \cite{schwinger1951gauge}, \cite{christensen1978regularization})
\begin{eqnarray}
&& \left\langle x,\mu;\tau |x',\nu;0\right\rangle ^{(1)} \sim  \frac{i}{16 \pi^2} 
|\mathrm{g}|^{\tfrac{1}{4}}\sqrt{\Delta}|\mathrm{g'}|^{\tfrac{1}{4}}e^{\tfrac{i\sigma}{2\tau}} 
\sum_{n=0} ^{\infty} (i\tau)^{n-2} b_{n \; \mu\nu'},  
\label{tr_ampl_expansion}\\
&&\lim_{x' \to x} b_{0 \; \mu\nu'} = g_{\mu\nu}(x), 
\label{init_cond_b0}
\end{eqnarray}
where $\Delta$ is the Van-Vleck Morette determinant $$\Delta \equiv -|{\rm g}| ^{-1/2}  
{\rm det } (\sigma_{\mu\nu'})  |{\rm g'}| ^{-1/2}$$ and
the coefficient bivectors $b_{n \; \mu\nu'}$ are evaluated by solving a recursion formula 
obtained upon insertion of \eqref{tr_ampl_expansion} into eq. \eqref{tr_ampl_eq}; 
such a recursion formula reads
\begin{eqnarray}
&& \sigma^{;\lambda}b_{0 \; \mu\nu' ;\lambda} = 0, \label{rec_rel_b0}\\
&&\sigma^{;\lambda}b_{n+1 \; \mu\nu';\lambda} + (n+1)b_{n+1 \; \mu\nu'} \nonumber \\
&& \;\;\;\;\;= \frac{1}{\sqrt{\Delta}}\left( \sqrt{\Delta}b_{n \; \mu\nu'}\right)
\tensor{}{_{;\lambda}^\lambda} - R^{\lambda}_\mu b_{n \; \lambda\nu'}.
\label{rec_rel_bn}
\end{eqnarray}
The equations \eqref{init_cond_b0} and \eqref{rec_rel_b0} are solved by $b_{0 \; \mu\nu' } 
= g_{\mu\nu'}$, which is known as the geodetic parallel displacement bi-vector; it is 
characterized by the following fundamental property: by applying it to a local contravariant 
vector $V^{\mu '}$ at $x'$ one obtains the covariant form of the vector which results from 
displacing $V^{\mu '}$ in a parallel fashion along the geodesic from $x'$ to $x$: 
$$  g_{\mu\nu'}V^{\nu '} = V_\mu . $$

It can be shown (see Ref. \cite{bimonte2003photon} for the detailed calculations) that as $x'$ 
approaches $x$ (which implies $\sigma(x,x') \to 0$), eqs. \eqref{reg_ph_Gr}, 
\eqref{tr_ampl_expansion}, \eqref{init_cond_b0} lead to
\begin{equation}
G_{\mu\nu'}^{(\xi)} \sim \frac{i}{16 \pi^2} \lim_{s \to 0} \frac{\mu_A ^{2s}}{\Gamma(s+1)} 
\mathcal{G}_{\mu\nu'} ^{(\xi)}(s), 
\label{Had_im}
\end{equation}
where, after having defined 
\begin{eqnarray}
\tensor{U}{_{n \;} _\mu ^\lambda}(s;\xi) &\equiv & \frac{2}{\sigma(x,x')} 
\underline{\delta}_{\mu} ^\lambda + \frac{(\xi^{s+1}-1)}{(s+n)(s+1)}
\nabla_\mu \nabla^\lambda, \\
\tensor{B}{_{n \;} _\lambda _{\nu'}}(s) &\equiv & \tensor{b}{_{n \;} _\lambda _{\nu'}} 
\sqrt{\Delta} (x,x') (\sigma (x,x') /2)^{s+n},
\end{eqnarray}
we write
\begin{equation}
\mathcal{G}_{\mu\nu'} ^{(\xi)}(s) \equiv \sum_{n=0} ^\infty \Gamma(1-s-n) 
\tensor{U}{_{n \;} _\mu ^\lambda}(s;\xi)\tensor{B}{_{n \;} _\lambda _{\nu'}}(s).
\label{inf_sum}
\end{equation}

What is crucial for us is the $s \to 0$ limit of the sum in the previous equation. 
Indeed, on studying first, for simplicity, the case when the gauge-field
operator reduces to a minimal (wavelike) operator, i.e., at $\xi = 1$, one finds
\begin{equation}
\mathcal{G}_{\mu\nu'} ^{(1)}(s) = \frac{2\sqrt{\Delta}(x,x')}{\sigma(x,x')} 
\sum_{n=0} ^{\infty} \tensor{f}{_{n \;} _\mu _{\nu'}}(s), 
\label{inf_sum_minimal}
\end{equation} 
having defined
\begin{equation}
\tensor{f}{_{n \;} _\mu _{\nu'}}(s) \equiv \Gamma(1-s-n) 
\tensor{b}{_{n \;} _\mu _{\nu'}} (\sigma(x,x')/2)^{s+n}.
\end{equation}
Since $\tensor{b}{_{0 \;} _\mu _{\nu'}} = \tensor{g}{ _\mu _{\nu'}}$, we therefore find
\begin{equation}
\mathcal{G}_{\mu\nu'} ^{(1)}(0) = \frac{2\sqrt{\Delta}(x,x')}{\sigma(x,x')}\tensor{g}{ _\mu _{\nu'}} 
+ \frac{2\sqrt{\Delta}(x,x')}{\sigma(x,x')}\lim_{s \to 0} \sum_{n=1}^{\infty} 
\tensor{f}{_{n \;} _\mu _{\nu'}}(s),
\end{equation}
where the fi\-rs\-t te\-rm on the rhs is pr\-ec\-is\-el\-y th\-e fi\-rs\-t te\-rm in t\-he 
Ha\-da\-ma\-rd as\-ym\-pt\-ot\-ic ex\-pa\-ns\-io\-n at sma\-ll $\sigma(x,x')$ (see Ref. 
\cite{christensen1978regularization} ). On the other hand, the Hadamard Green function is, 
apart from a factor $2$, the imaginary part of the Feynman Green function, in agreement with 
formula \eqref{Had_im}. Eventually, we find therefore, at small $\sigma(x,x')$,
\begin{eqnarray}
G_{\mu\nu'}^{(\xi)} &\sim & \frac{1}{8\pi^2}\frac{\sqrt{\Delta}(x,x')}{\sigma(x,x')
+i\epsilon}g_{\mu\nu'} + \frac{i}{16\pi^2} \lim_{s\to 0} \bigg[\frac{(\xi-1)}{s(s+1)}
\nabla_\mu \nabla^\lambda B_{0\;\lambda\nu'}(s) 
\nonumber \\
&&+ \sum_{n=1} ^{\infty} \Gamma(1-s-n) \tensor{U}{_{n \;} _\mu ^\lambda}(s;\xi)
\tensor{B}{_{n \;} _\lambda _{\nu'}}(s) \bigg],
\end{eqnarray}
i.e., the \lq\lq flat'' Feynman propagator, with the $i\epsilon$ term restored, plus 
correction resulting from the gauge parameter ($\xi \neq 1$ leading to a non-minimal 
operator) and from  the non-vanishing curvature.

It can be checked (see Ref. \cite{bimonte2003photon}) that this infinite sum \eqref{inf_sum} 
is also able to recover the familiar $\log \sigma (x,x')$ singularity, which occurs 
for massive theories in flat space-time and, more generally, even for massless 
theories (as in our case) but in curved space-time. 

\section{Second Derivatives, Stress-Energy Tensor}
\setcounter{equation}{0}

In this section it will be shown that, given a field theory (in the present case, quantum 
Maxwell's theory), there is a close relation between second derivatives of the Hadamard 
Green's functions, stress-energy tensor and effective action, following the footsteps of 
Christensen and DeWitt (see Refs. \cite{christensen1978regularization}, \cite{dewitt1975quantum1}). 
In a classical field theory, the stress-energy tensor can be obtained as
\begin{equation}
T^{\mu\nu} = \frac{2}{\sqrt{|\mathrm{g}}|} \frac{\delta S}{\delta g_{\mu\nu}}
\end{equation}
For a free theory, this object is quadratic in the fields; when dealing with the associated 
quantum theory, the expectation value of this observable is, in general, divergent: this 
happens because the two field operators are taken at the same space-time point. An useful 
way to regularize the expectation value of the stress-energy tensor is to insert into the 
formal expression for $T^{\mu\nu}$ not the field operators themselves but operators that 
have been smeared out by means of a smooth function $s$ of compact support:
\begin{equation}
\phi_s (x) \equiv \int d^4 y \;  s(x-y) \phi(y).
\end{equation}
The resulting operator does exist and the behaviour of its (finite) expectation value may be 
studied as the size of the support of $s(x)$ tends to zero. A regularization method equivalent to 
the one of the smearing method but easier to apply in practice is simply \textit{to separate the points} 
at which the two fiels in $T^{\mu\nu}$ are taken and then to 
\textit{examine the tensor as the points are brought together again}. 

For the action functional \eqref{full_ed_act}, the stress-energy tensor is
\begin{equation}
T^{\mu\nu} = \frac{2}{\sqrt{|\mathrm{g}}|} \frac{\delta \bar{S}}{\delta g_{\mu\nu}} 
= T^{\mu\nu} _{{\rm Maxwell}}+\tfrac{1}{\xi}T^{\mu\nu} _{{\rm gauge}}
+\tfrac{1}{\sqrt{\xi}}T^{\mu\nu} _{{\rm ghost}}, 
\label{starting_eq}
\end{equation}
where
\begin{eqnarray}
T^{\mu\nu} _{{\rm Maxwell}} &\equiv & \frac{\delta S_{{\rm Maxwell}}}{\delta g_{\mu\nu}} 
\equiv \frac{\delta}{\delta g_{\mu\nu}} \int d^4 x \; \left(-\frac{1}{4}|\mathrm{g}(x)|^{1/2}
F_{\mu\nu}F^{\mu\nu} \right),\\
T^{\mu\nu} _{{\rm gauge}} &\equiv & \frac{\delta S_{{\rm gauge}}}{\delta g_{\mu\nu}} 
\equiv \frac{\delta}{\delta g_{\mu\nu}}  \int d^4 x \; \left(-\frac{1}{2}|\mathrm{g}(x)|^{1/2}
\left(\tensor{A}{^\mu _{;\mu}}\right)^2 \right),\\
T^{\mu\nu} _{{\rm ghost}} &\equiv & \frac{\delta S_{{\rm ghost}}}{\delta g_{\mu\nu}} 
\equiv \frac{\delta}{\delta g_{\mu\nu}}  \int d^4 x \;\left( -|\mathrm{g}(x)|^{1/2}
\chi \Box \psi \right).
\end{eqnarray}
The result is 
\begin{eqnarray}
T^{\mu\nu} _{{\rm Maxwell}} &=& -\tensor{F}{^\mu _\rho}\tensor{F}{^\rho ^\nu} -\tfrac{1}{4}
F_{\alpha\beta}F^{\alpha\beta}  g^{\mu\nu},
\label{final_eq1}\\
T^{\mu\nu} _{{\rm gauge}} &=&  \left( \tfrac{1}{2}\left(\tensor{A}{^\alpha _{;\alpha}}\right)^2  
+\tensor{A}{^\alpha _{;\alpha\beta}} \tensor{A}{^\beta} \right)g^{\mu\nu} 
- \left(  \tensor{A}{_\alpha ^{;\alpha\mu}}\tensor{A}{^\nu} + \tensor{A}{_\alpha ^{;\alpha\nu}}
\tensor{A}{^\mu} \right),\label{final_eq2}\\
T^{\mu\nu} _{{\rm ghost}} &= & \chi_{;\alpha}  \psi^{;\alpha}g^{\mu\nu} 
- \left( \chi^{;\mu}\psi^{;\nu}+ \chi^{;\nu}\psi^{;\mu}  \right). 
\label{final_eq3}
\end{eqnarray}
The detailed calculations are provided in Appendix A.
In order to exploit the point split method, we write
\begin{eqnarray}
&& A_{\mu;\alpha} A_{\rho;\sigma} \nonumber \\
&=& \tfrac{1}{2} A_{\mu;\alpha} A_{\rho;\sigma} +  \tfrac{1}{2} A_{\rho;\sigma}A_{\mu;\alpha}  
\nonumber \\
&=& \tfrac{1}{2} \left[ A_{\mu;\alpha}, A_{\rho;\sigma} \right]_{+}  
\nonumber \\
&=& \lim_{x' \to x} \left\lbrace \tfrac{1}{4} \left[A_{\mu;\alpha}, A_{\rho';\sigma' }\right]_+ 
+  \tfrac{1}{4} \left[A_{\mu';\alpha'},A_{\rho;\sigma}\right] _+ \right\rbrace, 
\label{point_split}
\end{eqnarray}
where $[\; , \; ]_{+}$ is the anti-commutator. On the last line, we can pass from classical fields 
to quantum operators; define the matrix element of an operator $\Theta$ between 
$\langle{\rm out, vac}|$ and $|{\rm in, vac} \rangle$ as 
\begin{equation}
\langle \Theta \rangle ^{{\rm matrix}} \equiv \frac{\langle{\rm out, vac}|\Theta|
{\rm in, vac} \rangle }{\langle{\rm out, vac}|{\rm in, vac} \rangle}.
\end{equation}
Then, evaluating \eqref{point_split} between $\langle{\rm out, vac}|$ and $|{\rm in, vac} \rangle$, 
dividing by $$\langle{\rm out, vac}|{\rm in, vac} \rangle ,$$ and recalling that, up to a 
numerical factor, the matrix element of the anticommutator function is the Hadamard Green's 
function, one obtains\footnote{It should be pointed out that this is an abuse of notation: 
given for example two $2$-point tensors $B$, $C$, whose components are $B_{\mu\nu'}$, 
$C_{\mu\nu'}$, respectively, one can only sum the components pertaining to the same point 
and characterized by the same indices: in fact $$B_{\mu\nu'}+C_{\nu\mu'}$$ does not 
define a $2$-point tensor; therefore, throughtout the following, $$\lim_{x' \to x} 
\left\lbrace B_{\mu\nu'}+C_{\nu\mu'}\right\rbrace$$ stands for $$\left(\lim_{x' \to x}B_{\mu\nu'}\right) 
+ \left(\lim_{x' \to x}C_{\nu\mu'}\right)$$ which is, of course, a tensor at $x$.}
\begin{eqnarray}
A_{\mu;\alpha} A_{\rho;\sigma} \mapsto \lim_{x' \to x} \left\lbrace \tfrac{1}{4} 
G^{(H)}_{\mu\rho';\alpha\sigma'}+\tfrac{1}{4} G^{(H)}_{\rho\mu';\sigma\alpha'} \right\rbrace, 
\end{eqnarray}
where, in order to avoid any confusion with the Feynman Green function in the minimal case 
($\xi$=1), the Hadamard Green function shall be named $G^{(H)}_{\mu\nu'}$. In a similar manner
\begin{eqnarray}
&& A_{\mu;\nu\rho} A_{\sigma} = \tfrac{1}{4} [A_{\mu;\nu\rho} , A_{\sigma} ]_{+} 
+  \tfrac{1}{4} [A_{\mu';\nu'\rho'} , A_{\sigma} ]_{+} 
\nonumber \\
&& \; \; \mapsto \lim_{x' \to x} \left\lbrace \tfrac{1}{4} G^{(H)}_{\mu\sigma';\nu\rho} 
+  \tfrac{1}{4} G^{(H)}_{\sigma\mu';\nu'\rho'}\right\rbrace
\end{eqnarray}
and 
\begin{equation}
\chi_{;\mu}\psi_{;\nu} \mapsto \lim_{x' \to x} \left\lbrace \tfrac{1}{4} 
G^{(H)}_{;\mu\nu'}+\tfrac{1}{4} G^{(H)}_{;\nu\mu'}\right\rbrace,
\end{equation}
where 
$G^{(H)}(x,x')$ is the Hadamard Green's function for the ghost fields.

Hence the point-split method yields
\begin{eqnarray}
&&\langle T^{\alpha\beta} \rangle_{{\rm Maxwell}} ^{{\rm matrix}}= \tfrac{1}{4}\lim_{x' \to x}
\bigg\lbrace g^{\mu\sigma}(g^{\alpha\gamma}g^{\beta\rho}-\tfrac{1}{4}
g^{\alpha\beta}g^{\gamma\rho}) \bullet 
\nonumber \\
&& \; \; \; \; \bullet \; (   G^{(H)}_{\sigma\mu';\rho\gamma'} + G^{(H)}_{\mu\sigma';\gamma\rho'} 
+ G^{(H)}_{\rho\gamma';\sigma\mu'} + G^{(H)}_{\gamma\rho';\mu\sigma'} 
\nonumber \\
&& \; \; \; \; - G^{(H)}_{\rho\mu';\sigma\gamma'} - G^{(H)}_{\mu\rho';\gamma\sigma'} 
-G^{(H)}_{\sigma\gamma';\rho\mu'} - G^{(H)}_{\gamma\sigma';\mu\rho'}  ) \bigg\rbrace , 
\label{stress1}\\
&& \langle T^{\alpha\beta} \rangle_{{\rm gauge}} ^{{\rm matrix}} = \lim_{x' \to x}
\bigg\lbrace -\tfrac{1}{4}g^{\mu\rho}E^{\alpha\beta\nu\sigma}(G^{(H)}_{\rho\sigma';\mu\nu}
+G^{(H)}_{\rho\sigma';\mu'\nu'} )\nonumber \\
&& \; \; \; \; +\tfrac{1}{8}g^{\alpha\beta}g^{\mu\rho}g^{\nu\sigma}  
(G^{(H)}_{\rho\sigma';\mu\nu'}+G^{(H)}_{\sigma\rho';\nu\mu'} )\bigg\rbrace ,
\label{stress2}\\
&&\langle T^{\alpha\beta} \rangle_{{\rm ghost}} ^{{\rm matrix}} = -\tfrac{1}{4} 
\lim_{x' \to x}\bigg\lbrace E^{\alpha\beta\mu\nu}( G^{(H)}_{;\mu\nu'}
+ G^{(H)}_{;\nu\mu'})\bigg\rbrace, 
\label{stress3}
\end{eqnarray}
where the DeWitt supermetric has been introduced:
\begin{equation}
E^{\mu\nu\rho\tau} \equiv g^{\mu\rho}g^{\nu\tau} + g^{\mu\tau}g^{\nu\rho}-g^{\mu\nu}g^{\rho\tau}.
\end{equation}
We should now specify in which order the various operations we rely upon are performed. 
Indeed, in the evaluation of the Feynman Green function, we first sum over $n$ and then 
take the $s \to 0$ limit. Here, we eventually obtain the energy-momentum tensor of the 
quantum theory according to the point-splitting procedure, with the understanding 
that the coincidence limit $\lim_{x' \to x}$ is the last operation to be performed.

It is clear from \eqref{stress1}, \eqref{stress2}, \eqref{stress3} that our analysis of 
the stress-energy tensor is virtually completed if we can provide a closed expression for 
the coincidence limit of the second derivatives of the Hadamard Green function. 
It is easy to see that divergences appear; in the minimal case ($\xi$ = 1), 
on denoting by $\varepsilon$ the regulator that we are employing, the divergent 
part of $G^{(H)}_{\gamma\beta';\rho\tau'}$ is of the form (see Ref. \cite{bimonte2003photon})
\begin{equation}
\lim_{\varepsilon \to 0} \left\lbrace \Gamma(\varepsilon) ( [\tensor{b}{_{1 \;} _\gamma _{\beta';\rho\tau'}}]
-\tfrac{1}{6}[\tensor{b}{_{1 \;} _\gamma _{\beta'}}]R_{\rho\tau}) 
- \tfrac{1}{2}\Gamma(\varepsilon-1)[\tensor{b}{_{2 \;} _\gamma _{\beta'}}]g_{\rho\tau} \right\rbrace , 
\label{div_part}
\end{equation}
with the convention that the coincidence limit $\lim_{x' \to x}$ has to be taken 
for the quantities in square brackets.

In many applications, we are interested in finding
\begin{equation}
\langle T^{\mu\nu} \rangle^{{\rm vac}} \equiv \langle {\rm in, vac} 
|T^{\mu\nu}|{\rm in, vac} \rangle
\end{equation}
the vacuum expectation value of the stress-energy tensor in the vacuum state defined prior to 
any dynamics in the background gravitational field. This quantity, properly regularized and 
renormalized, gives us all the information we want about particle production and vacuum 
polarization. It is the object we choose to use as source in the semiclassical gravitational field equations,
\begin{equation}
G_{\mu\nu} = \langle T_{\mu\nu} \rangle^{{\rm vac}}, \label{back_reaction_grav}
\end{equation} 
when doing a back-reaction problem. So why are $\langle T^{\mu\nu} \rangle^{{\rm matrix}}$, 
the Green functions and their divergences interesting?
The answer is the following: DeWitt showed (see Ref. \cite{dewitt1975quantum1}) that
\begin{equation}
\langle T^{\mu\nu} \rangle^{{\rm vac}} =  \langle T^{\mu\nu} \rangle^{{\rm matrix}} 
+ \langle T^{\mu\nu} \rangle^{{\rm finite}},
\end{equation}
where $\langle T^{\mu\nu} \rangle^{{\rm finite}}$ vanishes when there is 
no particle annihilation, is always finite, and satisfies the conservation equation 
$\langle T^{\mu\nu} \rangle^{{\rm finite}} _{;\nu} = 0$ (the proof of this beautiful result 
is described in Appendix B). Hence the divergences appearing in $\langle T^{\mu\nu} 
\rangle^{{\rm vac}}$ and $\langle T^{\mu\nu} \rangle^{{\rm matrix}}$ are identical. 
Regularization of $\langle T^{\mu\nu} \rangle^{{\rm matrix}}$ yields the regularized form 
of $\langle T^{\mu\nu} \rangle^{{\rm vac}}$. Regularizing $\langle T^{\mu\nu} 
\rangle^{{\rm matrix}}$ gives:
\begin{equation}
\langle T^{\mu\nu} \rangle^{{\rm matrix}} = \langle T^{\mu\nu} \rangle^{{\rm div}} 
+ \langle T^{\mu\nu} \rangle^{{\rm matrix,ren}},
\end{equation}
where $\langle T^{\mu\nu} \rangle^{{\rm div}}$ contains the infinite pieces which we will 
renormalize away by adding infinite counterterms onto the classical action for the gravitational 
field and $\langle T^{\mu\nu} \rangle^{{\rm matrix,ren}}$ is the remaining finite physical 
part of the matrix element. Renormalizing $\langle T^{\mu\nu} \rangle^{{\rm div}}$ away 
also gives us a renormalized $\langle T^{\mu\nu} \rangle^{{\rm vac}}$
\begin{eqnarray}
\langle T^{\mu\nu} \rangle^{{\rm vac,ren}} &=& \langle T^{\mu\nu} \rangle^{{\rm vac}} 
- \langle T^{\mu\nu} \rangle^{{\rm div}} 
\nonumber \\
&=& \langle T^{\mu\nu} \rangle^{{\rm matrix,ren}} + \langle T^{\mu\nu} \rangle^{{\rm finite}},
\end{eqnarray}
to be used as the source in \eqref{back_reaction_grav}.

Another important fact is the close link between Feynman (and Hadamard) Green's functions 
and the so-called effective action $W$: DeWitt showed that
\begin{equation}
\langle T^{\mu\nu} \rangle^{{\rm matrix}} = 2 \mathrm{|g|}^{-1/2} \frac{\delta W_{}}{\delta g_{\mu\nu}},
\end{equation}
where 
\begin{equation}
W = -i \log \langle {\rm out,vac} |{\rm in, vac} \rangle.
\end{equation}
Having defined 
\begin{equation}
W = \int d^4 x \; L_{{\rm eff}},
\end{equation}
he also found that
\begin{equation}
L_{{\rm eff}} = {\rm Im} \lim_{x' \to x} {\rm tr} \frac{\partial}{\partial \sigma} 
\left(\mathrm{|g|}^{1/4}(x) G(x,x')   \mathrm{|g|}^{1/4}(x')  \right).
\end{equation}
Thus the renormalized effective Lagrangian is
\begin{equation}
L_{{\rm eff,ren}} = L_{{\rm eff}} - L_{{\rm div}},
\end{equation} 
where $L_{{\rm div}}$ can be evaluated by means of the divergent part of the asymptotic 
expansion of Feynman and Hadamard Green's functions.

Interestingly, the divergent part of the one-loop effective action for the quantum version 
of Einstein gravity has been recently discussed in Ref. \cite{giacchini2020} in relation to 
renormalization group equations for the Newton constant and the cosmological constant; the 
reader may find there an up-todate discussion of the concepts just introduced in our section.

\section{Divergent part of $\langle T^{\mu\nu} \rangle$}
\setcounter{equation}{0}

In this section we will show the results of the calculation pertaining to the divergent part 
of the coincidence limit of the Hadamard Green function and their application to the evaluation 
of the divergent part of $\langle T^{\mu\nu} \rangle$; the divergent part of the 
Hadamard Green function is (see \eqref{div_part})
\begin{equation}
\left[ G^{(H)}_{\gamma\beta';\rho\tau'} \right]^{\rm div} =\lim_{\varepsilon \to 0} \left\lbrace 
\Gamma(\varepsilon) ( [\tensor{b}{_{1 \;} _\gamma _{\beta';\rho\tau'}}]-\tfrac{1}{6}
[\tensor{b}{_{1 \;} _\gamma _{\beta'}}]R_{\rho\tau}) - \tfrac{1}{2}\Gamma(\varepsilon-1)
[\tensor{b}{_{2 \;} _\gamma _{\beta'}}]g_{\rho\tau} \right\rbrace .
\end{equation}
Thus, upon using 
\begin{equation}
\Gamma(\varepsilon -k) = \frac{1}{\varepsilon}\frac{(-1)^k}{k!} + O(1), \;\;\; {\rm for} 
\; k=0, 1, 2, ... \Longrightarrow
\begin{cases}
\Gamma(\varepsilon) =  \frac{1}{\varepsilon} + O(1), \\
\Gamma(\varepsilon -1) = - \frac{1}{\varepsilon} + O(1)
\end{cases}
\end{equation}
together with coincidence limits for the coefficient bivectors and for the remaining 
geometrical quantities which appear in the Fock-Schwinger-DeWitt asymptotic expansion 
(most of them are carefully derived in Appendices C, D) one obtains
\begin{eqnarray}
\varepsilon \left[ G^{(H)}_{\gamma\beta';\rho\tau'} \right]^{\rm div} &=&  
[g_{\beta\gamma}\;g_{\rho\tau}(-\tfrac{1}{360}\tensor{R}{_{\alpha_{1}} _{\alpha_{2}}} 
\tensor{R}{^{\alpha_{1}} ^{\alpha_{2}}} + \tfrac{1}{144}R^2 
\nonumber \\
&&+ \tfrac{1}{360}\tensor{R}{_{\alpha_{1}} _{\alpha_{2}} _{\alpha_{3}} _{\alpha_{4}}} 
\tensor{R}{^{\alpha_{1}} ^{\alpha_{2}} ^{\alpha_{3}} ^{\alpha_{4}}} + \tfrac{1}{60}\Box R)
\nonumber \\
&&+g_{\beta\gamma}(- \tfrac{1}{135}R_{{\alpha_1}{\alpha_2}{\alpha_3}\rho} 
\tensor{R}{^{{\alpha_1}{\alpha_2}{\alpha_3}} _\tau}- \tfrac{1}{135}R_{{\alpha_1}{\alpha_2}{\alpha_3}\rho} 
\tensor{R}{^{{\alpha_1}{\alpha_3}{\alpha_2}} _\tau}
\nonumber \\
&&+ \tfrac{1}{45}R_{{\alpha_1}\rho} R^{\alpha_1}_\tau   + \tfrac{1}{40}R_{;\rho\tau} 
- \tfrac{1}{60}\Box R_{\rho\tau}- \tfrac{1}{36}R\; R_{\rho\tau}
\nonumber \\
&&- \tfrac{1}{90}R^{{\alpha_1} {\alpha_2}} R_{{\alpha_2}\rho{\alpha_1}\tau} 
+ \tfrac{1}{20}\tensor{R}{_\rho ^{\alpha_1} _; _{\alpha_1} _\tau}                               
- \tfrac{1}{30}\tensor{R}{_\tau ^{\alpha_1} _; _{\alpha_1} _\rho} )    
\nonumber \\          
&& +g_{\rho\tau}( \tfrac{1}{4}R_{{\alpha_1}\beta} R^{\alpha_1}_\gamma - \tfrac{1}{12}
\Box R_{\beta\gamma}- \tfrac{1}{12}R\; R_{\beta\gamma} 
\nonumber \\
&& -\tfrac{1}{24}R_{{\alpha_1}{\alpha_2}{\alpha_3}\beta} \tensor{R}{^{{\alpha_1}{\alpha_2}
{\alpha_3}} _\gamma}) + \tfrac{1}{6}R_{\beta\gamma}\;R_{\rho\tau} 
\nonumber \\
&& + \tfrac{1}{12}\tensor{R}{^{\alpha_1} _\gamma _{\alpha_2} _\tau} \tensor{R}
{^{\alpha_2} _\rho _{\alpha_1} _\beta } 
+ \tfrac{1}{12}\tensor{R}{^{\alpha_1} _\gamma _{\alpha_2} _\rho} 
\tensor{R}{^{\alpha_2} _\tau _{\alpha_1} _\beta } 
\nonumber \\ 
&& + \tfrac{1}{12}\tensor{R}{^{\alpha_1} _\rho  _\beta _\gamma  _; _{\alpha_1} _\tau} 
- \tfrac{1}{12}\tensor{R}{^{\alpha_1} _\tau _\beta _\gamma  _; _{\alpha_1} _\rho} 
+ \tfrac{1}{3}R_{\beta} ^{\alpha_1} R_{{\alpha_1}\gamma\rho\tau}
\nonumber \\ 
&& - \tfrac{1}{6}R_{\gamma} ^{\alpha_1} R_{{\alpha_1}\beta\rho\tau}  
- \tfrac{1}{6}R_{\beta\gamma;\rho\tau}  
- \tfrac{1}{12}R \; R_{\beta\gamma\rho\tau}].
\end{eqnarray}
In order to obtain the divergent part of $\langle T^{\mu\nu} \rangle$, one also needs 
\begin{itemize}
\item[1.] The so-called \textit{quantum Ward identities}: they constitute a relation 
between gauge-field Green's 
functions and ghost Green's function. In our case, they read
\begin{eqnarray}
-G^{(H)}_{;\mu} &=& \tensor{G}{^{(H)} _{\mu\nu ' ;} ^{\nu '}}, \\
-G^{(H)}_{;\nu '} &=& \tensor{G}{^{(H)} _{\mu\nu ' ;} ^{\mu}}.
\end{eqnarray}
Therefore, by taking another covariant derivative, one obtains
\begin{eqnarray}
-G^{(H)}_{;\mu\rho'} &=& \tensor{G}{^{(H)} _{\mu\nu' ;} ^{\nu'} _{\rho'}}, \\
-G^{(H)}_{;\nu' \rho} &=& \tensor{G}{^{(H)} _{\mu\nu' ;} ^{\mu} _{\rho}}.
\end{eqnarray}
	
\item[2.] An expression for the divergent part of the coincidence limit of 
$G^{(H)}_{\rho\sigma';\mu\nu}$ and $G^{(H)}_{\rho\sigma';\mu'\nu'}$ in terms of 
$G^{(H)}_{\rho\sigma';\mu\nu'}$. They are easily obtained by using the parallel displacement matrix:
\begin{eqnarray}
\left[G^{(H)}_{\rho\sigma';\mu\nu}\right] &=& 
\left[G^{(H)}_{\rho\sigma';\mu\beta'} \tensor{g}{^{\beta'} _\nu}\right] 
\nonumber \\
&=& \left[G^{(H)}_{\rho\sigma';\mu\beta'}\right]\left[\tensor{g}{^{\beta'} _\nu}\right] 
\nonumber \\
&=& \left[G^{(H)}_{\rho\sigma';\mu\beta'}\right] \underline{\delta}^{\beta} _\nu 
\nonumber \\
&=& \left[G^{(H)}_{\rho\sigma';\mu\nu'}\right], \\
&& \nonumber \\
\left[G^{(H)}_{\rho\sigma';\mu'\nu'}\right] &=& 
\left[G^{(H)}_{\rho\sigma';\beta\nu'}\tensor{g}{^\beta _{\mu'}}\right] 
\nonumber \\
&=& \left[G^{(H)}_{\rho\sigma';\beta\nu'}\right]\left[ \tensor{g}{^\beta_{\mu'}}\right] 
\nonumber \\
&=& \left[G^{(H)}_{\rho\sigma';\beta\nu'}\right] \underline{\delta}^{\beta} _\mu 
\nonumber \\
&=& \left[G^{(H)}_{\rho\sigma';\mu\nu'}\right].
\end{eqnarray}
\end{itemize}
The final result is therefore
\begin{eqnarray}
\varepsilon \langle T^{\mu\nu} \rangle^{{\rm div}} &=&   
[g^{\mu\nu} ( -\tfrac{23}{180} R_{\alpha\beta} R^{\alpha\beta} 
+\tfrac{1}{144} R^2 - \tfrac{1}{24} R_{\alpha\beta\gamma\delta} R^{\alpha\gamma\beta\delta} 
\nonumber \\
&&\; \; \; \; \; -\tfrac{7}{180} R_{\alpha\beta\gamma\delta}R^{\alpha\beta\gamma\delta} 
+ \tfrac{7}{120} \Box R  ) 
\nonumber \\
&& +\tfrac{17}{180} R_{\alpha\beta}R^{\alpha\mu\nu\beta} +\tfrac{127}{1080} 
\tensor{R}{^{\mu\alpha\beta\gamma}} \tensor{R}{^\nu _{\alpha\beta\gamma}} 
\nonumber \\
&& -\tfrac{49}{540} \tensor{R}{^{\mu\alpha\beta\gamma}} \tensor{R}{^\nu _{\beta\alpha\gamma}}
+\tfrac{19}{180}\tensor{R}{^\mu _{\alpha}}R^{\alpha\nu} 
\nonumber \\
&& +\tfrac{1}{20}(\tensor{R}{^\mu ^\alpha _; _\alpha ^\nu}
+\tensor{R}{^\nu ^\alpha _; _\alpha ^\mu}) 
\nonumber \\
&& +\tfrac{1}{24}(\tensor{R}{ ^\alpha ^\mu _; ^\nu _\alpha }
+\tensor{R}{ ^\alpha ^\nu _; ^\mu _\alpha }) 
\nonumber \\
&& -\tfrac{7}{120}R^{;\mu\nu}+\tfrac{1}{18} RR^{\mu\nu} -\tfrac{11}{60}\Box R^{\mu\nu}]. 
\label{Tstress_def}
\end{eqnarray}
Although the expressions for the divergent part of $\left[ G^{(H)}_{\gamma\beta';\rho\tau'} \right]$ 
and $\langle T^{\mu\nu} \rangle^{{\rm div}}$ may seem cumbersome, there is a systematic way 
to deal with them. Indeed, a more careful inspection shows that they result from a very large 
number of terms: by looking at the coincidence limits in Appendices C, D, it is easy to see that 
each divergent part of $\left[ G^{(H)}_{\gamma\beta';\rho\tau'} \right]$ is obtained from the 
sum of eighty contractions variously involving the metric tensor, the Ricci and Riemann tensor, 
and that $\langle T^{\mu\nu} \rangle^{{\rm div}}$ is obtained by summing thirty of these objects; 
therefore expression \eqref{Tstress_def} for $\langle T^{\mu\nu} \rangle^{{\rm div}}$ results 
from more than two thousand terms. For this purpose, a program has been written in FORM.
In the following two sections, eq. \eqref{Tstress_def} will be made more explicit in some 
physically relevant particular cases.

\section{Ricci-flat space-times}
\setcounter{equation}{0}

In light of their role in General Relativity, Ricci-flat metric tensors are very relevant, 
since they solve the Einstein vacuum equations. Whenever one considers a Ricci-flat metric, 
only four terms are left in the sum eq. \eqref{Tstress_def}, i.e., the ones which 
only involve the Riemann tensor and its derivatives:
\begin{eqnarray}
\varepsilon \langle T^{\mu\nu} \rangle^{{\rm div}}\bigg |_{\rm Ricci-flat} &=&  
[g^{\mu\nu} ( - \tfrac{1}{24} R_{\alpha\beta\gamma\delta} R^{\alpha\gamma\beta\delta} 
-\tfrac{7}{180} R_{\alpha\beta\gamma\delta}R^{\alpha\beta\gamma\delta}  ) 
\nonumber \\
&&  +\tfrac{127}{1080} \tensor{R}{^{\mu\alpha\beta\gamma}} \tensor{R}{^\nu _{\alpha\beta\gamma}} 
-\tfrac{49}{540} \tensor{R}{^{\mu\alpha\beta\gamma}} \tensor{R}{^\nu _{\beta\alpha\gamma}}]. 
\label{stresstensorRicci}
\end{eqnarray}
An important example of vacuum solution is the Kerr metric, which describes the exterior of a 
rotating, stationary, axially symmetric star; it depends on two parameters: the 
Schwarzschild radius $r_s \equiv \frac{2GM}{c^2}$ and the angular momentum $J$, through 
$a \equiv \frac{J}{Mc}$; in a spherical coordinate system $(t,r,\theta,\phi)$, the metric 
components $g_{\mu\nu}$ form the matrix 
\begin{equation} 
g_{\mu\nu}=\left(
\begin{array}{cccc}
g_{00} & 0 & 0 & g_{03} \\
0 & g_{11} & 0 & 0 \\
0 & 0 & g_{22}  & 0 \\
g_{30} & 0 & 0 & g_{33} \\
\end{array}
\right),
\end{equation}
where
\begin{eqnarray}
&&g_{00} = \frac{r r_s}{r^2+a^2 \cos ^2(\theta )}-1, 
\nonumber \\
&&g_{11} = \frac{r^2+a^2 \cos ^2(\theta )}{a^2+r^2-r r_s}, 
\nonumber \\
&&g_{22} = r^2+a^2 \cos ^2(\theta ), 
\nonumber \\
&&g_{33} = \sin ^2(\theta ) \left(\frac{r r_s \sin ^2(\theta ) a^2}
{r^2+a^2 \cos ^2(\theta )}+a^2+r^2\right), 
\nonumber \\
&&g_{03}=g_{30} = \frac{a r r_s \sin ^2(\theta )}{r^2+a^2 \cos ^2(\theta )} 
\nonumber .
\end{eqnarray}
Then eq. \eqref{stresstensorRicci} is 
\begin{eqnarray}
\varepsilon \langle T^{\mu\nu} \rangle^{{\rm div}} \bigg |_{\rm Kerr}=\left(
\begin{array}{cccc}
\mathcal{A}^{00} & 0 & 0 & \mathcal{A}^{03} \\
0 & \mathcal{A}^{11} & 0 & 0 \\
0 & 0 &\mathcal{A}^{22} & 0 \\
\mathcal{A}^{30}& 0 & 0 & \mathcal{A}^{33} \\
\end{array}
\right),
\end{eqnarray}
where 
\begin{eqnarray}
&&\mathcal{A}^{00}= -\frac{ \mathcal{Y}r_s ^2 (a^4 + 2r^4 +a^2 r (3 r +r_s)+a^2(a^2+r(r -r_s)
\cos (2\theta) ) }{\mathcal{X}\mathcal{Z}}, 
\nonumber  \\
&&\mathcal{A}^{11}=\frac{2\mathcal{X}\mathcal{Y} r_s ^2 }{\mathcal{Z}},  
\nonumber \\
&&\mathcal{A}^{22}=\frac{2 \mathcal{Y}  r_s ^2}{\mathcal{Z}} ,   
\nonumber\\
&&\mathcal{A}^{33}= \frac{\mathcal{Y}r_s^2 (a^2+2r(r-r_s)
+a^2\cos (2\theta))\csc(\theta)^2}{\mathcal{X}\mathcal{Z}},  
\nonumber\\
&&\mathcal{A}^{03}=\mathcal{A}^{30} = \frac{2\mathcal{Y} a r r_s ^3}{\mathcal{X}\mathcal{Z}} ,  
\nonumber
\end{eqnarray}
having defined 
\begin{eqnarray}
&&\mathcal{X}=a^2 + r(r-r_s),  
\nonumber \\
&&\mathcal{Y}\equiv a^6 \cos (6 \theta )+10 a^6-180 a^4 r^2+240 a^2 r^4
+6 a^4 \left(a^2-10 r^2\right) \cos (4 \theta ) 
\nonumber \\
&&\;\;\;\;\;\;\;+15 a^2 \left(a^4-16 a^2 r^2+16 r^4\right) \cos (2 \theta )-32 r^6,  
\nonumber \\
&&\mathcal{Z}\equiv (a^2 +2r^2 +a^2 \cos (2\theta))^7  
\nonumber.
\end{eqnarray}

Of course, if $J=0$, then $a=0$, and the Kerr metric reduces to the Schwarzschild metric; 
in this case, the same calculations yield
\begin{equation}
\varepsilon \langle T^{\mu\nu} \rangle^{{\rm div}}\bigg |_{\rm Schwarzschild}=
\left(
\begin{array}{cccc}
\frac{r_s ^2}{2 r^6-2 r^5 r_s} & 0 & 0 & 0 \\
0 & \frac{r_s^2 (r_s-r)}{2 r^7} & 0 & 0 \\
0 & 0 & -\frac{r_s^2}{2 r^8} & 0 \\
0 & 0 & 0 & -\frac{r_s^2 \csc ^2(\theta )}{2 r^8} \\
\end{array}
\right).
\end{equation}

\section{Maximally symmetric spaces}
\setcounter{equation}{0}

Modern cosmology is based on the hypothesis that on a large enough scale the universe is 
spatially homogeneous and isotropic. Together, these two assumptions are known as the 
\textit{cosmological principle}. As is well known, spaces which are both spatially homogenous 
and isotropic are maximally symmetric, i.e., they possess the largest possible number 
of Killing vector fields which in a $n$-dimensional manifold equals $n(n+1)/2$. It can be shown 
that the following holds for a maximally symmetric space:
\begin{itemize}
\item[1.] The scalar curvature $R$ is a constant;
\item[2.] The Ricci tensor is proportional to the metric tensor, i.e., $$R_{\mu\nu} 
= \tfrac{1}{n} R g_{\mu\nu};$$
\item[3.] The Riemann curvature tensor is given by $$R_{\mu\nu\lambda\rho}=\tfrac{R}{n(n-1)} 
(g_{\mu\lambda}g_{\nu\rho}-g_{\nu\lambda}g_{\mu\rho}).$$
\end{itemize}
As a consequence, all terms which involve derivatives in eq. \eqref{Tstress_def} vanish; 
therefore only nine terms are left in the sum, and eq. \eqref{Tstress_def} 
takes the simplified form 
\begin{eqnarray}
\varepsilon \langle T^{\mu\nu} \rangle^{{\rm div}} &=&  
[g^{\mu\nu} ( -\tfrac{23}{180} R_{\alpha\beta} R^{\alpha\beta} +\tfrac{1}{144} 
R^2 - \tfrac{1}{24} R_{\alpha\beta\gamma\delta} R^{\alpha\gamma\beta\delta} 
\nonumber \\
&&\; \; \; \; \; -\tfrac{7}{180} R_{\alpha\beta\gamma\delta}R^{\alpha\beta\gamma\delta} ) 
\nonumber \\
&& +\tfrac{17}{180} R_{\alpha\beta}R^{\alpha\mu\nu\beta} +\tfrac{127}{1080} 
\tensor{R}{^{\mu\alpha\beta\gamma}} \tensor{R}{^\nu _{\alpha\beta\gamma}} 
\nonumber \\
&& -\tfrac{49}{540} \tensor{R}{^{\mu\alpha\beta\gamma}} \tensor{R}{^\nu _{\beta\alpha\gamma}}
+\tfrac{19}{180}\tensor{R}{^\mu _{\alpha}}R^{\alpha\nu} 
\nonumber \\
&& +\tfrac{1}{18} RR^{\mu\nu}]. 
\label{Tstressmaxsymm}
\end{eqnarray}
A crucial example is de Sitter space, which is the maximally symmetric vacuum solution of 
Einstein's field equations with a positive cosmological constant $\Lambda$ (corresponding 
to a positive vacuum energy density and negative pressure), actually; the cosmological 
constant is linked to the Hubble constant by $$H = \left(\frac{\Lambda}{3} \right)^{1/2}.$$ 
It is often useful to coordinatize the space-time in two different ways, depending upon 
whether one wishes to think of it as an expanding Friedmann-Lema{\^i}tre-Robertson-Walker 
or a static universe with an event horizon. In the former case $(s,\chi,\theta,\phi)$ are used, 
while $(t,r,\theta,\phi)$ coordinates are employed in the latter; the metric components thus become 
\begin{equation} 
g_{\mu\nu}= \left(
\begin{array}{cccc}
-1 & 0 & 0 & 0 \\
0 & \frac{\cosh ^2(H s)}{H^2} & 0 & 0 \\
0 & 0 & \frac{\cosh ^2(H s) \sin ^2(\chi )}{H^2} & 0 \\
0 & 0 & 0 & \frac{\cosh ^2(H s) \sin ^2(\theta ) \sin ^2(\chi )}{H^2} \\
\end{array}
\right),
\end{equation}
or 
\begin{equation} 
g_{\mu\nu}= \left(
\begin{array}{cccc}
H^2 r^2-1 & 0 & 0 & 0 \\
0 & \frac{1}{1-H^2 r^2} & 0 & 0 \\
0 & 0 & r^2 & 0 \\
0 & 0 & 0 & r^2 \sin ^2(\theta ) \\
\end{array}
\right)
\end{equation}

Hence, the components of the divergent part of the stress energy-tensor are, in the former case
\begin{equation}
\varepsilon \langle T^{\mu\nu} \rangle^{{\rm div}} \bigg |_{ {\rm de \; Sitter} \; 
(s,\chi,\theta,\phi)  } = 
\left(
\begin{array}{cccc}
\mathcal{B}_{00} & 0 & 0 & 0 \\
0 &\mathcal{B}_{11}  & 0 & 0 \\
0 & 0 &\mathcal{B}_{22}   & 0 \\
0 & 0 & 0 &\mathcal{B}_{33}   \\
\end{array}
\right),
\end{equation}
where
\begin{eqnarray}
&&\mathcal{B}_{00}=\tfrac{5}{2} H^4, 
\nonumber \\
&&\mathcal{B}_{11}= -\tfrac{5}{2}  H^6 \text{sech}^2(H s), 
\nonumber \\
&&\mathcal{B}_{22}=-\tfrac{5}{2} H^6 \csc ^2(\chi ) \text{sech}^2(H s), 
\nonumber \\
&&\mathcal{B}_{33}=-\tfrac{5}{2}  H^6 \csc ^2(\theta ) \csc ^2(\chi ) \text{sech}^2(H s), 
\nonumber \\
\end{eqnarray}
while in the latter 
\begin{equation}
\varepsilon \langle T^{\mu\nu} \rangle^{{\rm div}} \bigg |_{ {\rm de \; Sitter} \; 
(t,r,\theta,\phi)  } = 
\left(
\begin{array}{cccc}
\mathcal{C}_{00} & 0 & 0 & 0 \\
0 &\mathcal{C}_{11}  & 0 & 0 \\
0 & 0 &\mathcal{C}_{22}   & 0 \\
0 & 0 & 0 &\mathcal{C}_{33}   \\
\end{array}
\right),
\end{equation}
where
\begin{eqnarray}
&&\mathcal{C}_{00}=\frac{5H^4}{2-2H^2 r^2} , 
\nonumber \\
&&\mathcal{C}_{11}= \tfrac{5}{2}  H^4 (-1+H^2 r^2) , 
\nonumber \\
&&\mathcal{C}_{22}=-\frac{5 H^4}{2r^2} , 
\nonumber \\
&&\mathcal{C}_{33}=-\frac{5H^4 \csc ^2(\theta )}{2r^2} . 
\nonumber \\
\end{eqnarray}

\section{Concluding remarks}

In this paper the Feynman propagator for Maxwell's theory in curved space-time has been 
described by means of the Fock-Schwinger-DeWitt asymptotic expansion; its crucial role has been emphasized, 
together with the point-splitting method, in the evaluation and regularization of quadratic 
(in the dynamical field) observables. Among these, there is the stress-energy tensor: its matrix 
element has been derived in terms of second covariant derivatives of the Hadamard Green function of the 
electromagnetic field. Remarkably, the divergences occurring in $\langle T^{\mu\nu} \rangle^{{\rm vac}}$ 
and $\langle T^{\mu\nu} \rangle^{{\rm matrix}}$ are identical, therefore regularizing 
$\langle T^{\mu\nu} \rangle^{{\rm matrix}}$ is 
equivalent to regularizing $\langle T^{\mu\nu} \rangle^{{\rm vac}}$. 
An original computation has been then presented: a concise, explicit formula for the divergent 
part of the stress-energy tensor. This has been obtained from a careful handling of more than two thousand 
terms, and in order to perform such a calculation, a program has been written in FORM 
(see Refs. \cite{Heck1}, \cite{Vermaseren1}, \cite{courses1}). Such a formula holds for every 
space-time, and is fully evaluated for some relevant examples: Kerr and Schwarzschild metrics 
(which are Ricci-flat) and de Sitter metric (which is maximally symmetric); see Refs. 
\cite{teukolsky2015kerrmetric}, \cite{realdi2008einstein} for an up-todate review.
As far as we know, our original results in Eqs. (4.10), (5.3), (5.4), (6.4) and (6.6) 
are not available in the literature (cf. Refs. \cite{Brown}, \cite{Jensen}).

In any theory of interacting fields the set of currents that describe
the interaction is of fundamental importance; in General Relativity, these currents are the components 
of the stress-energy tensor, therefore \textit{the main problem in developing a quantum 
field theory in curved space-time is precisely to understand the
stress-energy tensor} (see Ref. \cite{dewitt1975quantum1}). Compared to the flat space-time case, in
a curved background the resulting renormalized stress-energy tensor is covariantly conserved, 
of course, but it possesses a state-independent anomalous trace (see Refs. 
\cite{christensendoct1975}, \cite{duff1977}, \cite{christensenduff1978}, \cite{kay2006}).

The results obtained are interesting also in the context of effective action theory in curved space-time, 
whose divergent part is essential to discuss renormalization group equations for the Newton constant 
and the cosmological constant \cite{giacchini2020}. 
Moreover, they can be used to obtain a proper source in the semiclassical 
gravitational field equations, when doing a back reaction problem, for every background gravitational field.

\section*{Acknowledgments}

R. N. would like to acknowledge the helpful advice that he received
from S. M. Christensen.
G. E.  is grateful to the ``Ettore Pancini'' Physics
Department of Federico II University for hospitality and support.

\begin{appendix}

\section{{$ T^{\mu\nu}$ Calculations}}
\setcounter{equation}{0}

In this appendix the calculations regarding the derivation of \eqref{starting_eq} from 
\eqref{final_eq1}, \eqref{final_eq2}, \eqref{final_eq3} are explicitly shown: \\ \\
$T^{\mu\nu} _{{\rm Maxwell}}$ calculations: 
\begin{eqnarray}
&&\delta ( -\tfrac{1}{4}|\mathrm{g}|^{1/2}F_{\mu\nu}F^{\mu\nu} )  
\nonumber \\
&=& \; \; -\tfrac{1}{4}F_{\mu\nu}F_{\rho\sigma}(g^{\mu\rho})(g^{\nu\sigma}) (\delta |\mathrm{g}|^{1/2}) 
\nonumber \\
&& \; \; -\tfrac{1}{4}F_{\mu\nu}F_{\rho\sigma}(\delta g^{\mu\rho})(g^{\nu\sigma})(|\mathrm{g}|^{1/2}) 
\nonumber \\
&& \; \; -\tfrac{1}{4}F_{\mu\nu}F_{\rho\sigma}(g^{\mu\rho})(\delta g^{\nu\sigma})(|\mathrm{g}|^{1/2}) 
\nonumber \\
&=& -\tfrac{1}{4}F_{\mu\nu}F_{\rho\sigma}(g^{\mu\rho})(g^{\nu\sigma})(\tfrac{1}{2}
|\mathrm{g}|^{1/2} g^{\alpha\beta}\delta g_{\alpha\beta})  
\nonumber \\
&& \; \; -\tfrac{1}{4}F_{\mu\nu}F_{\rho\sigma}(-g^{\mu\alpha}g^{\rho\beta}\delta g_{\alpha\beta })
(g^{\nu\sigma})(|\mathrm{g}|^{1/2}) 
\nonumber \\
&& \; \; -\tfrac{1}{4}F_{\mu\nu}F_{\rho\sigma}(g^{\mu\rho})(- g^{\nu\alpha}g^{\sigma\beta}
\delta g_{\alpha\beta})(|\mathrm{g}|^{1/2}) 
\nonumber \\
&=& -\tfrac{1}{8}F_{\mu\nu}F^{\mu\nu}|\mathrm{g}|^{1/2}g^{\alpha\beta}\delta g_{\alpha\beta}  
\nonumber \\
&& \; \; -\tfrac{1}{2}\tensor{F}{^\alpha _\rho}\tensor{F}{^\rho ^\beta} 
|\mathrm{g}|^{1/2}\delta g_{\alpha\beta}.
\end{eqnarray}
Therefore 
\begin{equation}
T^{\mu\nu} _{{\rm Maxwell}} = -\tensor{F}{^\mu _\rho}\tensor{F}{^\rho ^\nu} 
-\tfrac{1}{4}F_{\alpha\beta}F^{\alpha\beta}  g^{\mu\nu}.
\end{equation}
$T^{\mu\nu} _{{\rm gauge}}$ calculations:
\begin{eqnarray}
&&\delta  \left(-\tfrac{1}{2}|\mathrm{g}(x)|^{1/2}\left(\tensor{A}{^\mu _{;\mu}}\right)^2 \right)  
\nonumber \\
&=& \; \; -\tfrac{1}{2}(\delta |\mathrm{g}(x)|^{1/2})\left(\tensor{A}{^\mu _{;\mu}}\right)^2 
\nonumber \\
&& \; \; -( |\mathrm{g}(x)|^{1/2})\tensor{A}{^\eta _{;\eta}}\left[\delta (g^{\mu\nu}\nabla_\nu A_\mu) \right] 
\nonumber \\
&=& -\tfrac{1}{2}(\tfrac{1}{2}|\mathrm{g}(x)|^{1/2}g^{\alpha\beta}\delta g_{\alpha\beta})
\left(\tensor{A}{^\mu _{;\mu}}\right)^2   
\nonumber \\ 
&& \; \; -( |\mathrm{g}(x)|^{1/2})\tensor{A}{^\eta _{;\eta}} \left[\delta g^{\mu\nu}
\nabla_\nu A_\mu + g^{\mu\nu}\delta (\partial_\nu A_\mu - \Gamma^{\rho}_{\mu\nu}A_\rho)\right] 
\nonumber \\
&=& -\tfrac{1}{4}|\mathrm{g}(x)|^{1/2}\left(\tensor{A}{^\mu _{;\mu}}\right)^2  
g^{\alpha\beta}\delta g_{\alpha\beta} 
\nonumber \\ 
&& \; \; +|\mathrm{g}(x)|^{1/2}\tensor{A}{^\eta _{;\eta}} g^{\mu\alpha}g^{\nu\beta}
\delta g_{\alpha\beta}\nabla_\nu A_\mu 
\nonumber \\
&& \; \; + |\mathrm{g}(x)|^{1/2}\tensor{A}{^\eta _{;\eta}} 
g^{\mu\nu}\delta \Gamma^{\rho}_{\mu\nu}A_\rho 
\nonumber \\
&=& -\tfrac{1}{4}|\mathrm{g}(x)|^{1/2}\left(\tensor{A}{^\mu _{;\mu}}\right)^2  
g^{\alpha\beta}\delta g_{\alpha\beta} 
\nonumber \\ 
&& \; \; + |\mathrm{g}(x)|^{1/2}\tensor{A}{^\eta _{;\eta}} 
\delta g_{\alpha\beta} A^{\alpha ; \beta} 
\nonumber \\
&& \; \; + |\mathrm{g}(x)|^{1/2}\tensor{A}{^\eta _{;\eta}} g^{\mu\nu} \left[\tfrac{1}{2}g^{\rho\alpha}
(\delta g_{\alpha\nu;\mu} +  \delta g_{\alpha\mu;\nu} - \delta g_{\mu\nu;\alpha}) \right]A_\rho 
\nonumber \\
&=& -\tfrac{1}{4}|\mathrm{g}(x)|^{1/2}\left(\tensor{A}{^\mu _{;\mu}}\right)^2  
g^{\alpha\beta}\delta g_{\alpha\beta} 
\nonumber \\ 
&& \; \; + |\mathrm{g}(x)|^{1/2}\tensor{A}{^\eta _{;\eta}} 
\delta g_{\alpha\beta} A^{\alpha ; \beta} 
\nonumber \\
&& \; \; + |\mathrm{g}(x)|^{1/2}\tensor{A}{^\eta _{;\eta}} g^{\mu\nu}A^{\alpha}\delta 
g_{\alpha\nu;\mu} -\tfrac{1}{2}|\mathrm{g}(x)|^{1/2}
\tensor{A}{^\eta _{;\eta}}A^{\alpha}g^{\mu\nu}\delta g_{\mu\nu;\alpha}. 
\nonumber
\end{eqnarray}
Integrating by parts on the last line, one obtains
\begin{eqnarray}
&&\delta  \left(-\tfrac{1}{2}|\mathrm{g}(x)|^{1/2}
\left(\tensor{A}{^\mu _{;\mu}}\right)^2 \right)  
\nonumber \\
&=& -\tfrac{1}{4}|\mathrm{g}(x)|^{1/2}\left(\tensor{A}{^\mu _{;\mu}}\right)^2 
g^{\alpha\beta}\delta g_{\alpha\beta} 
\nonumber \\ 
&& \; \; + |\mathrm{g}(x)|^{1/2}\tensor{A}{^\eta _{;\eta}}  
A^{\alpha ; \beta}\delta g_{\alpha\beta} 
\nonumber \\
&& \; \; - |\mathrm{g}(x)|^{1/2}\tensor{A}{^\eta _{;\eta\mu}}g^{\mu\nu}\tensor{A}{^\alpha}
\delta g_{\alpha\nu} -  |\mathrm{g}(x)|^{1/2}\tensor{A}{^\eta _{;\eta}}g^{\mu\nu}
\tensor{A}{^{\alpha} _{;\mu}}\delta g_{\alpha\nu} 
\nonumber \\
&& \; \; + \tfrac{1}{2}|\mathrm{g}(x)|^{1/2}\tensor{A}{^\eta _{;\eta\alpha}}
\tensor{A}{^\alpha}g^{\mu\nu}\delta g_{\mu\nu}+ \tfrac{1}{2}|\mathrm{g}(x)|^{1/2}
\tensor{A}{^\eta _{;\eta}}\tensor{A}{^\alpha _{;\alpha}}g^{\mu\nu}\delta g_{\mu\nu} 
\nonumber \\
&=& \left( \tfrac{1}{4}\left(\tensor{A}{^\mu _{;\mu}}\right)^2  +\tfrac{1}{2}
\tensor{A}{^\mu _{;\nu\rho}} \tensor{A}{^\rho} \right)|\mathrm{g}(x)|^{1/2}
g^{\alpha\beta}\delta g_{\alpha\beta} 
\nonumber \\
&& \; \; - \tfrac{1}{2} \left(  \tensor{A}{_\eta ^{;\eta\alpha}}\tensor{A}{^\beta} 
+ \tensor{A}{_\eta ^{;\eta\beta}}\tensor{A}{^\alpha} \right) 
|\mathrm{g}(x)|^{1/2}\delta g_{\alpha\beta}. 
\end{eqnarray}
Therefore 
\begin{equation}
T^{\mu\nu} _{{\rm gauge}} =  \left( \tfrac{1}{2}\left(\tensor{A}{^\alpha _{;\alpha}}\right)^2  
+\tensor{A}{^\alpha _{;\alpha\beta}} \tensor{A}{^\beta} \right)g^{\mu\nu} 
-  \left(  \tensor{A}{_\alpha ^{;\alpha\mu}}\tensor{A}{^\nu} 
+ \tensor{A}{_\alpha ^{;\alpha\nu}}\tensor{A}{^\mu} \right).
\end{equation}
$T^{\mu\nu} _{{\rm ghost}}$ calculations:
\begin{eqnarray}
&&\delta(-|\mathrm{g}(x)|^{1/2}\chi \Box \psi) 
\nonumber \\
&=& -\delta(|\mathrm{g}(x)|^{1/2})\chi \Box \psi -|\mathrm{g}(x)|^{1/2}\chi \delta(\Box \psi) 
\nonumber \\
&=& -\tfrac{1}{2}|\mathrm{g}(x)|^{1/2})\chi \Box \psi g^{\mu\nu}\delta g_{\mu\nu} 
\nonumber \\
&& \; \; -|\mathrm{g}(x)|^{1/2}\chi \delta \left( g^{\nu\mu} \partial _\mu \partial_\nu \psi 
- g^{\nu\mu} \Gamma^{\rho}_{\mu\nu}\partial_\rho \psi\right) 
\nonumber \\
&=&  -\tfrac{1}{2}|\mathrm{g}(x)|^{1/2}\chi \Box \psi g^{\mu\nu}\delta g_{\mu\nu} 
\nonumber \\
&& \; \; +|\mathrm{g}(x)|^{1/2}\chi g^{\nu\alpha}g^{\mu\beta}\delta  g_{\alpha\beta} 
\partial _\mu \partial _\nu \psi 
\nonumber \\
&& \; \; -|\mathrm{g}(x)|^{1/2}\chi g^{\nu\alpha}g^{\mu\beta}\delta  g_{\alpha\beta} 
\Gamma^{\rho}_{\mu\nu}\partial_\rho \psi 
\nonumber \\
&& \; \; +|\mathrm{g}(x)|^{1/2}\chi g^{\nu\mu} \delta \Gamma^{\rho}_{\mu\nu}\partial_\rho \psi 
\nonumber \\
&=&  -\tfrac{1}{2}|\mathrm{g}(x)|^{1/2}\chi \Box \psi g^{\mu\nu}\delta g_{\mu\nu} 
\nonumber \\
&& \; \; +|\mathrm{g}(x)|^{1/2}\chi g^{\nu\alpha}g^{\mu\beta}\delta  g_{\alpha\beta}  \psi_{;\nu\mu} 
\nonumber \\
&& \; \; +|\mathrm{g}(x)|^{1/2}\chi g^{\nu\mu} (\tfrac{1}{2}g^{\rho\eta}(\nabla_\mu 
\delta g_{\nu\eta}+\nabla_\nu \delta g_{\mu\eta}-\nabla_\eta \delta g_{\mu\nu}))\partial_\rho \psi 
\nonumber \\
&=&  -\tfrac{1}{2}|\mathrm{g}(x)|^{1/2}\chi \Box \psi g^{\mu\nu}\delta g_{\mu\nu} 
\nonumber \\
&& \; \; +|\mathrm{g}(x)|^{1/2}\chi g^{\nu\alpha}g^{\mu\beta}\delta  g_{\alpha\beta} \psi_{;\nu\mu}  
\nonumber \\
&& \; \; +|\mathrm{g}(x)|^{1/2}\chi g^{\nu\mu} g^{\rho\eta}\delta g_{\nu\eta;\mu} \partial_\rho \psi  
-\tfrac{1}{2}|\mathrm{g}(x)|^{1/2}\chi g^{\nu\mu} g^{\rho\eta}\delta g_{\mu\nu;\eta}\partial_\rho \psi . 
\nonumber 
\end{eqnarray}
Upon integrating by parts on the last line, one obtains
\begin{eqnarray}
&=&  -\tfrac{1}{2}|\mathrm{g}(x)|^{1/2}\chi \Box \psi g^{\mu\nu}\delta g_{\mu\nu} 
\nonumber \\
&& \; \; +|\mathrm{g}(x)|^{1/2}\chi g^{\nu\alpha}g^{\mu\beta}\delta  g_{\alpha\beta} \psi_{;\nu\mu} 
\nonumber \\
&& \; \; -|\mathrm{g}(x)|^{1/2}\chi g^{\nu\mu} g^{\rho\eta}\delta g_{\nu\eta}\psi_{;\rho\mu} 
-|\mathrm{g}(x)|^{1/2}\chi_{;\mu} g^{\nu\mu} g^{\rho\eta}\delta g_{\nu\eta} \psi_{;\rho} 
\nonumber \\
&& \; \; +\tfrac{1}{2}|\mathrm{g}(x)|^{1/2}\chi_{;\eta} g^{\nu\mu} g^{\rho\eta}\delta g_{\mu\nu} 
\psi_{;\rho} +\tfrac{1}{2}|\mathrm{g}(x)|^{1/2}\chi g^{\nu\mu} 
g^{\rho\eta}\delta g_{\mu\nu} \psi_{;\rho\eta} 
\nonumber \\
&=&  -\tfrac{1}{2}|\mathrm{g}(x)|^{1/2}\chi \tensor{\psi}{_{;\alpha}^\alpha} 
g^{\mu\nu}\delta g_{\mu\nu} 
\nonumber \\
&& \; \; +|\mathrm{g}(x)|^{1/2}\chi \tensor{\psi}{^{;\alpha\beta}}\delta  g_{\alpha\beta}  
\nonumber \\
&& \; \; -|\mathrm{g}(x)|^{1/2}\chi \psi^{;\eta\nu} \delta g_{\nu\eta} 
-|\mathrm{g}(x)|^{1/2}\chi^{;\mu}\psi^{;\eta}\delta g_{\mu\eta} 
\nonumber \\
&& \; \; +\tfrac{1}{2}|\mathrm{g}(x)|^{1/2}\chi_{;\eta}  \psi^{;\eta}g^{\nu\mu} 
\delta g_{\mu\nu} +\tfrac{1}{2}|\mathrm{g}(x)|^{1/2}\chi g^{\nu\mu} 
\tensor{\psi}{_{;\eta}^\eta} \delta g_{\mu\nu} 
\nonumber \\
&=& \tfrac{1}{2}|\mathrm{g}(x)|^{1/2}\chi_{;\alpha}  \psi^{;\alpha}g^{\mu\nu} 
\delta g_{\mu\nu} -\tfrac{1}{2}|\mathrm{g}(x)|^{1/2}\left(\chi^{;\mu}\psi^{;\nu}
+ \chi^{;\nu}\psi^{;\mu}\right)\delta g_{\mu\nu}. 
\nonumber 
\end{eqnarray}
Therefore
\begin{equation}
T^{\mu\nu} _{{\rm ghost}} = \chi_{;\alpha}  \psi^{;\alpha}g^{\mu\nu} 
- \left( \chi^{;\mu}\psi^{;\nu}+ \chi^{;\nu}\psi^{;\mu}  \right).
\end{equation}

\section{DeWitt's beautiful result}
\setcounter{equation}{0}
\subsection{\lq\lq In'' and \lq\lq Out'' regions, Bogoliubov Coefficients}

In this appendix the proof of the following fundamental result will be outlined: 
\begin{equation}
\langle T^{\mu\nu} \rangle^{{\rm vac}} =  \langle T^{\mu\nu} \rangle^{{\rm matrix}} 
+ \langle T^{\mu\nu} \rangle^{{\rm finite}},
\end{equation}
where $\langle T^{\mu\nu} \rangle^{{\rm finite}}$ vanishes when there is no particle 
annihilation, is always finite, and satisfies the conservation equation 
$\langle T^{\mu\nu} \rangle^{{\rm finite}} _{;\nu} = 0$.

Let us assume that space-time has two causally connected stationary regions, an \lq\lq in'' 
region and an \lq\lq out'' region, each possessing complete Cauchy hypersurfaces and a 
timelike Killing vector, which makes it possible to define positive and negative 
frequencies. Let us consider a free theory, i.e., a theory whose action fuctional $S$ 
is quadratic in the (real) dynamical field; then the equation of motion is given in terms 
of a linear, self-adjoint, differential operator $F$, to which a conserved inner product on 
the space of solutions is associated. Hence, given a complete set $\lbrace u_i \rbrace$ of 
normalised basis solutions which contain only positive frequencies, 
the field solution can be expanded in the form
\begin{eqnarray}
\phi(x) = \sum_i \left(a_i u_i(x) + a_i ^\dagger u_i ^*(x) \right). 
\end{eqnarray}
By using the canonical (anti)commutation relations, or, in a more elegant and manifestly 
covariant way, by using the Peierls \cite{peierls1997commutation} definition of 
(anti-)commutator, it is easy to see that the operator coefficients in the expansion satisfy 
the (anti-)commutation relations
\begin{eqnarray}
&&[a_i , a_j ^\dagger]_\pm = \underline{\delta}_{ij}, \\
&&[a_i , a_j ]_\pm = 0.
\end{eqnarray}
This operator algebra serves in the traditional fashion to define a Fock space and a \lq\lq vacuum'' state:
\begin{equation}
a_i |{\rm vac}\rangle = 0.
\end{equation}
In our case, there are two sets $\lbrace u _{i \; \rm in}\rbrace$, $\lbrace u _{i \; \rm out}\rbrace$ 
of normalised basis solutions which contain only positive frequencies, associated with the 
\lq\lq in'' and \lq\lq out'' regions, respectively; they will be connected by a Bogoliubov transformation
\begin{equation}
u_{i \; \rm out} = \sum_j \left(\alpha_{ij}u_{j \; \rm in} + \beta_{ij}u_{j \; \rm in}^* \right);
\end{equation} 
then we have
\begin{equation}
\phi(x) = \sum_i \left(a_{i \; \rm in} u_{i \; \rm in}(x) 
+ a_{i \; \rm in} ^\dagger u_{i \; \rm in} ^* (x)  
\right) = \sum_i \left(a_{i \; \rm out} u_{i \; \rm out}(x) 
+ a_{i \; \rm out} ^\dagger u_{i \; \rm out} ^* (x)  \right) 
\end{equation}
and the \lq\lq vacuum'' state vectors are defined by
\begin{eqnarray}
&&a_{i \; \rm in} |{\rm in, vac}\rangle = 0, \\
&&a_{i \; \rm out} |{\rm out, vac}\rangle = 0.
\end{eqnarray}
The annihilation operators in the \lq\lq in'' and \lq\lq out'' regions are related by
\begin{eqnarray}
&& a_{i \; \rm out} = \sum_j \left(\alpha_{ij}^* a_{j \; \rm in} 
- \beta_{ij}^* a_{j \; \rm in}^\dagger \right), \\
&& a_{i \; \rm in} = \sum_j \left(\alpha_{ji} a_{j \; \rm out} 
- \beta_{ji}^* a_{j \; \rm out}^\dagger \right).
\end{eqnarray}

\subsection{Creation and Annihilation Amplitudes, Vacuum-to-Vacuum Amplitude, 
Relations of its divergencies with those of $T_{\mu\nu}$}

Of particular importance in quantum cosmology are the many-particle production and annihilation amplitudes:
\begin{eqnarray}
&& i^{n/2}V_{i_1...i_n} \equiv e^{-iW}\langle {\rm out}, i_1...i_n|{\rm in, vac}\rangle , \\
&& i^{n/2}\Lambda_{i_1...i_n} \equiv e^{-iW}\langle {\rm out, vac}|{\rm in},i_1...i_n\rangle ,
\end{eqnarray}
where
\begin{eqnarray}
&&e^{iW} \equiv \langle {\rm out, vac}|{\rm in, vac}\rangle , \\
&& |{\rm in},i_1...i_n\rangle \equiv  a_{i_1 \; \rm in}^\dagger ...  
a_{i_n \; \rm in}^\dagger |{\rm in, vac}\rangle , \\
&& |{\rm out},i_1...i_n\rangle \equiv  a_{i_1 \; \rm out}^\dagger ...  
a_{i_n \; \rm out}^\dagger |{\rm out, vac}\rangle .
\end{eqnarray}
Assuming unit normalization for the \lq\lq in'' and \lq\lq out'' vacuum state vectors, one may write
\begin{eqnarray}
&& |{\rm in, vac}\rangle = e^{iW} \sum_{n=0}^{+\infty} \frac{i^{n/2}}{n!} 
\sum_{j_1 ... j_n} V_{j_1...j_n} |{\rm out},j_1...j_n\rangle , \\
&& |{\rm out, vac}\rangle = e^{-iW^*} \sum_{n=0}^{+\infty} \frac{(-i)^{n/2}}{n!} 
\sum_{j_1 ... j_n} \tensor{\Lambda}{_{j_1...j_n}^*} |{\rm in},j_1...j_n\rangle .
\end{eqnarray}
It can be easily proven that the production amplitudes $V_{i_1...i_n}$ and the 
annihilation amplitudes $\Lambda_{i_1...i_n}$ obey
\begin{eqnarray}
&& V_{i_1...i_n} = 
\begin{cases}
0 &  \; \;  n \; {\rm odd,} \\
\sum_p V_{i_1 i_2} \; ... \; V_{i_{n-1} i_n} & \; \; n  \; {\rm even,} \\
\end{cases} \\
&& \Lambda_{i_1...i_n} = 
\begin{cases}
0 &  \; \;  n \; {\rm odd,} \\
\sum_p \Lambda_{i_1 i_2} \; ... \; \Lambda_{i_{n-1} i_n} & \; \; n  \; {\rm even,} \\
\end{cases}
\end{eqnarray}
where $\sum_p$ denotes a summation over the $\tfrac{n!}{2^{n/2}(n/2)!} $ distinct pairings of the 
labels $i_1, ..., i_n$. The previous equations reveal that particle production and annihilation 
processes as composed of individual pair creation and annihilation events.

Consider now an infinitesimal change $\delta g_{\mu\nu}$ in the metric tensor; it yields a 
change $\delta S$ in the action functional for the field $\phi$; if the support of 
$\delta g_{\mu\nu}$ is confined to the space-time region between the \lq\lq in'' and 
\lq\lq out'' regions then, by Schwinger variational principle, it follows that:
\begin{eqnarray}
\delta W &=& -i e^{-iW}\delta e^{iW} = -i e^{-iW}\delta 
\langle {\rm out, vac}|{\rm in, vac}\rangle 
\nonumber \\ 
&=&  e^{-iW} \langle {\rm out, vac}|\delta S|{\rm in, vac}\rangle ,
\end{eqnarray}
which implies 
\begin{equation}
\frac{2}{\sqrt{|\mathrm{g}}|} \frac{\delta W}{\delta g_{\mu\nu}}= e^{-iW} 
\langle {\rm out, vac}|T^{\mu\nu}|{\rm in, vac}\rangle 
\equiv \langle T^{\mu\nu}\rangle ^{{\rm matrix}}. 
\label{DeW_result1}
\end{equation}
Introduce now the following notation, in which integration over space-time coordinates is suppressed:
\begin{eqnarray}
S[\phi]&=& \tfrac{1}{2} \phi F \phi, \\
\frac{\delta S}{\delta \phi}[\phi] &=& F \phi .
\end{eqnarray}
Then a more precise notation for the stress energy tensor can be introduced:
\begin{eqnarray}
&& T^{\mu\nu}(\chi,\psi) \equiv \chi \frac{\delta F}{\delta g_{\mu\nu}} \psi, \\
&&\frac{2}{\sqrt{|\mathrm{g}}|}\frac{\delta S}{\delta g_{\mu\nu}}[\phi] 
=  \phi \frac{\delta F}{\delta g_{\mu\nu}} \phi \equiv T^{\mu\nu}(\phi,\phi);
\end{eqnarray}
therefore eq. \eqref{DeW_result1} is
\begin{equation}
\langle T^{\mu\nu}\rangle ^{{\rm matrix}} = e^{-iW} 
\langle {\rm out, vac}|T^{\mu\nu}(\phi,\phi)|{\rm in, vac}\rangle . 
\end{equation}
Using the field expansion and the creation/annihilation amplitudes, one obtains
\begin{eqnarray}
&&\langle T^{\mu\nu}\rangle ^{{\rm matrix}}  
\nonumber \\
&=&  \sum_i T^{\mu\nu}(\tensor{u}{_{i \; \rm in}}, \tensor{u}{_{i \; \rm in}^*}) 
+ i \sum_{i,j} \Lambda_{ij}T^{\mu\nu}( \tensor{u}{_{i \; \rm in}^*}, 
\tensor{u}{_{j \; \rm in}^*});
\end{eqnarray}
The first term is
\begin{equation}
\sum_i T^{\mu\nu}(\tensor{u}{_{i \; \rm in}}, \tensor{u}{_{i \; \rm in}^*}) 
= \langle{\rm in, vac}|T^{\mu\nu}(\phi,\phi)|{\rm in, vac}\rangle 
= \langle T^{\mu\nu}(\phi,\phi)\rangle ^{{\rm vac}} \equiv 
\langle T^{\mu\nu}\rangle ^{{\rm vac}}, 
\end{equation}
while the second term is zero when there is no particle annihilation (because of the presence 
of $\Lambda_{ij}$) and is always finite; let us show that it satisfies the conservation 
equation; it suffices to note that
\begin{equation}
T^{\mu\nu}_{;\nu}(\phi,\phi) {\;\;\;\rm for \; every \; solution \;} \phi 
\label{Dew_result2}
\end{equation}
and that
\begin{eqnarray}
&&T^{\mu\nu}( \tensor{u}{_{i \; \rm in}^*}+\tensor{u}{_{j \; \rm in}^*}, 
\tensor{u}{_{i \; \rm in}^*}+\tensor{u}{_{j \; \rm in}^*}) 
\nonumber \\
&=& T^{\mu\nu}( \tensor{u}{_{i \; \rm in}^*}, \tensor{u}{_{i \; \rm in}^*})
+T^{\mu\nu}( \tensor{u}{_{j \; \rm in}^*}, \tensor{u}{_{j \; \rm in}^*})
+ 2 T^{\mu\nu}( \tensor{u}{_{i \; \rm in}^*}, \tensor{u}{_{j \; \rm in}^*});
\end{eqnarray}
therefore \eqref{Dew_result2} implies 
\begin{equation}
T^{\mu\nu}_{;\nu}( \tensor{u}{_{i \; \rm in}^*}, \tensor{u}{_{j \; \rm in}^*})=0,
\end{equation}
i.e., the conservation equation. Hence it is proven that 
\begin{equation}
\langle T^{\mu\nu} \rangle^{{\rm vac}} =  \langle T^{\mu\nu} \rangle^{{\rm matrix}} 
+ \langle T^{\mu\nu} \rangle^{{\rm finite}},
\end{equation}
where $\langle T^{\mu\nu} \rangle^{{\rm finite}}$ is zero when there is no particle 
annihilation, is always finite, and satisfies the conservation equation 
$\langle T^{\mu\nu} \rangle^{{\rm finite}} _{;\nu} = 0$; it can be expressed as
\begin{equation}
\langle T^{\mu\nu} \rangle^{{\rm finite}} = - i \sum_{i,j} \Lambda_{ij}T^{\mu\nu}
( \tensor{u}{_{i \; \rm in}^*}, \tensor{u}{_{j \; \rm in}^*}).
\end{equation}

\section{World Function: Coincidence Limits}
\subsection{General equations satisfied by the covariant derivatives of the world function}
\setcounter{equation}{0}

Following Synge in Ref. \cite{synge1931characteristic}, we will show a method to calculate 
the coincidence limits of the covariant derivatives of the world function, up to order six; 
these quantities are crucial in the evaluation of the divergent part of the stress-energy 
tensor. First, one needs to recall the fundamental equations satisfied by the world function:
\begin{eqnarray}
\sigma_{;\mu}\sigma^{;\mu} &=& 2 \sigma, 
\label{fund1} \\
\sigma_{;\mu\nu\rho} - \sigma_{;\mu\rho\nu} &=& 
- R_{\mu\lambda\nu\rho} \sigma^{;\lambda}. 
\label{fund2}
\end{eqnarray}
Differentiating the first equation, and indicating with a subscript that a certain index is free, one obtains
\begin{eqnarray}
&&\tensor{\sigma}{_{;\alpha_1} _{\mu} }\sigma^{;\mu} =  \sigma_{;\alpha_1}, 
\label{fund1_der} \\
&&\tensor{\sigma}{_{;\alpha_1 \mu\alpha_2} }\sigma^{;\mu}
+\tensor{\sigma}{_{;\alpha_1} _{\mu} }\tensor{\sigma}{^{;\mu}_{\alpha_2}} 
=  \sigma_{;\alpha_1 \alpha_2}, 
\label{fund1_der2}\\
&&\tensor{\sigma}{_{;\alpha_1 \mu\alpha_2 \alpha_3} }\sigma^{;\mu}
+\tensor{\sigma}{_{;\alpha_1 \mu\alpha_2} }\tensor{\sigma}{^{;\mu} _{\alpha_3}} 
+ \tensor{\sigma}{_{;\alpha_1} _{\mu} _{\alpha_3} }\tensor{\sigma}{^{;\mu}_{\alpha_2}}
+\tensor{\sigma}{_{;\alpha_1} _{\mu} }\tensor{\sigma}{^{;\mu}_{\alpha_2} _{\alpha_3}} 
\nonumber \\
&& \; \; \; \; \;= \sigma_{;\alpha_1 \alpha_2 \alpha_3}, 
\label{fund1_der3}
\end{eqnarray}
while differentiating the second one, one obtains
\begin{eqnarray}
&&\tensor{\sigma}{_{;{\alpha_1}{\alpha_2}{\alpha_3}{\alpha_4}}} 
- \tensor{\sigma}{_{;{\alpha_1}{\alpha_3}{\alpha_2}{\alpha_4}}} 
= - \tensor{R}{_{{\alpha_1}\lambda{\alpha_2}{\alpha_3};{\alpha_4}}} 
\tensor{\sigma}{^{;\lambda}}- \tensor{R}{_{{\alpha_1}\lambda{\alpha_2}{\alpha_3}}} 
\tensor{\sigma}{^{;\lambda}_{\alpha_4}}, 
\label{fund2_der}\\
&&\tensor{\sigma}{_{;{\alpha_1}{\alpha_2}{\alpha_3}{\alpha_4}{\alpha_5}}} 
- \tensor{\sigma}{_{;{\alpha_1}{\alpha_3}{\alpha_2}{\alpha_4}{\alpha_5}}}  
\nonumber \\
&& \;\;\;\;\;=- \tensor{R}{_{{\alpha_1}\lambda{\alpha_2}{\alpha_3};{\alpha_4}{\alpha_5}}} 
\tensor{\sigma}{^{;\lambda}}- \tensor{R}{_{{\alpha_1}\lambda{\alpha_2}{\alpha_3};{\alpha_4}}} 
\tensor{\sigma}{^{;\lambda}_{\alpha_5}} 
\nonumber \\
&& \;\;\;\;\;- \tensor{R}{_{{\alpha_1}\lambda{\alpha_2}{\alpha_3};{\alpha_5}}} 
\tensor{\sigma}{^{;\lambda}_{\alpha_4}}- \tensor{R}{_{{\alpha_1}\lambda{\alpha_2}{\alpha_3}}} 
\tensor{\sigma}{^{;\lambda} _{\alpha_4} _{\alpha_5}},   
\label{fund2_der2}\\
&&\tensor{\sigma}{_{;{\alpha_1}{\alpha_2}{\alpha_3}{\alpha_4}{\alpha_5}{\alpha_6}}} 
- \tensor{\sigma}{_{;{\alpha_1}{\alpha_3}{\alpha_2}{\alpha_4}{\alpha_5}{\alpha_6}}}  
\nonumber \\
&& \;\;\;\;\;=- \tensor{R}{_{{\alpha_1}\lambda{\alpha_2}{\alpha_3};{\alpha_4}{\alpha_5}{\alpha_6}}} 
\tensor{\sigma}{^{;\lambda}}- \tensor{R}{_{{\alpha_1}\lambda{\alpha_3}{\alpha_2};
{\alpha_4}{\alpha_5}}} \tensor{\sigma}{^{;\lambda} _{\alpha_6}}
\nonumber \\
&&\;\;\;\;\;- \tensor{R}{_{{\alpha_1}\lambda{\alpha_2}{\alpha_3};{\alpha_4}{\alpha_6}}} 
\tensor{\sigma}{^{;\lambda}_{\alpha_5}} 
- \tensor{R}{_{{\alpha_1}\lambda{\alpha_3}{\alpha_2};{\alpha_4}}}  
\tensor{\sigma}{^{;\lambda}_{\alpha_5} _{\alpha_6}} 
\nonumber \\
&& \;\;\;\;\; - \tensor{R}{_{{\alpha_1}\lambda{\alpha_2}{\alpha_3};{\alpha_5}{\alpha_6}}} 
\tensor{\sigma}{^{;\lambda}_{\alpha_4}}- \tensor{R}{_{{\alpha_1}\lambda{\alpha_2}{\alpha_3};{\alpha_5}}} 
\tensor{\sigma}{^{;\lambda}_{\alpha_4} _{\alpha_6}}
\nonumber \\
&&\;\;\;\;\;- \tensor{R}{_{{\alpha_1}\lambda{\alpha_2}{\alpha_3};{\alpha_6}}} 
\tensor{\sigma}{^{;\lambda} _{\alpha_4} _{\alpha_5}}- \tensor{R}{_{{\alpha_1}
\lambda{\alpha_2}{\alpha_3}}} \tensor{\sigma}{^{;\lambda} _{\alpha_4} _{\alpha_5} _{\alpha_6}}.  
\label{fund2_der3}
\end{eqnarray}
For the sake of readability, the following notation will be introduced: when an index 
is not a summation index, it will be indicated with a number, e.g. 
$$\tensor{\sigma}{_{;{\alpha_1}{\alpha_2}\lambda{\alpha_3}{\alpha_4}{\alpha_5}}} 
\equiv \tensor{\sigma}{_{;12\lambda 345}},$$ 
and the Riemann tensor, with its derivatives, will be indicated through the sequence 
of its indices between parentheses, e.g. 
$$ \tensor{R}{_{{\alpha_1}\lambda{\alpha_2}{\alpha_3};{\alpha_6}}} \equiv (1,\lambda ,2,3;6).$$
Then the equations which result from differentiations of \eqref{fund1} can be put in the form
\begin{equation}
\sigma_{;123...l} = \sum \tensor{\sigma}{_{;1\lambda {a_2}{a_3}...{a_k}}}
\tensor{\sigma}{^{;\lambda} _{a_{k+1}} _{...} _{a_l}} 
\label{fund1_gen}
\end{equation}
where the summation extends over all expressions for which $a_2 ,a_3,...,a_k$ is a set 
selected in order from $2,...,l$ and $a_{k+1}, ..., a_l$ is the remainder of the set 
$2,...,l$ in order, and $k$ takes all values from $2$ to $l$.
The equations which follow from differentiations on \eqref{fund2} take the form
\begin{equation}
\sigma_{;123...l}-\sigma_{;132...l} = - \sum (1,\lambda,2,3;b_4,...,b_k) 
\tensor{\sigma}{^{;\lambda} _{b_{k+1}} _{...} _{b_l}}, 
\label{fund2_gen}
\end{equation}
where the summation extends over all expressions for which $b_4,...,b_k$ is a set selected 
in order from $4,...,l$ and $b_{k+1},..., b_l$ is the remainder of the set 
$4,...,l$ in order, and $k$ takes all values from $4$ to $l$. 
\subsection{Coincidence limits of covariant derivatives}
Dividing \eqref{fund1_der} by $\sigma$, one obtains
\begin{equation}
\tensor{\sigma}{_{;1} _{\mu} }t^{\mu} =  t_{1},
\end{equation}  
where $t$ is the unit tangent vector along the geodesic; then the coincidence limit reads
\begin{equation}
[ \sigma_{;12} ] = g_{12}.
\end{equation}
From \eqref{fund1_der2} and \eqref{fund2_der} one has
\begin{equation}
[ \sigma_{;123} ] + [\sigma _{;132}] = 0 = [ \sigma_{;123} ] - [\sigma _{;132}],
\end{equation}
which implies
\begin{equation}
[ \sigma_{;123} ] =0.
\end{equation}
For the next order, by \eqref{fund1_der3} and \eqref{fund2_der2} one has
\begin{eqnarray}
&&[\sigma_{;1234} ]+[\sigma_{;1423} ]+[\sigma_{;1324} ] = 0, 
\label{limit4_1} \\
&&[\sigma_{;1234} ]-[\sigma_{;1323} ] = -(1,4,2,3). 
\label{limit4_2}
\end{eqnarray}
By \eqref{limit4_2} we convert \eqref{limit4_1} into
\begin{equation}
[\sigma_{;1243} ] + 2 [\sigma_{;1234} ] + (1,3,2,4) + (1,4,2,3) = 0; 
\label{limit4_3}
\end{equation}
Interchanging $3$ and $4$ and subtracting, we get
\begin{equation}
[\sigma_{;1234} ]= [\sigma_{;1243} ], 
\label{limit4_sym}
\end{equation}
and thus \eqref{limit4_3} gives 
\begin{equation}
[\sigma_{;1234} ] = -\tfrac{1}{3} ( (1,3,2,4)+(1,4,2,3) ),
\end{equation} 
or 
\begin{equation}
[\sigma_{;1234} ] = -\tfrac{1}{3} \mathcal{P}_{3,4} (1,3,2,4), 
\label{limit4_res}
\end{equation}
where $\mathcal{P}_{3,4}$ denotes the sum of expressions obtained by permutation of $3$, $4$.

For the fifth order, by \eqref{fund1_gen} and \eqref{fund2_gen},
\begin{eqnarray}
&& [\sigma_{;15234}]+[\sigma_{;14235}]+[\sigma_{;13245}]+[\sigma_{;12345}]=0, 
\label{limit5_1} \\
&& [\sigma_{;12345}]-[\sigma_{;13245}]=-(1,4,2,3;5)-(1,5,2,3;4). 
\label{limit5_2}
\end{eqnarray}
By \eqref{limit5_2} we convert \eqref{limit5_1} into
\begin{eqnarray}
&&[\sigma_{;12534}]+[\sigma_{;12435}]+2[\sigma_{;12345}] 
\nonumber \\
&&\;\;\;\;\;+(1,4,2,3;5)+(1,5,2,3;4)+(1,3,2,4;5)
\nonumber \\
&&\;\;\;\;\;+(1,5,2,4;3)+(1,3,2,5;4)+(1,4,2,5;3) =0. 
\label{limit5_3}
\end{eqnarray}
Interchanging $4$ and $5$, and subtracting, we get
\begin{equation}
[\sigma_{;12345}] = [\sigma_{;12354}] , 
\label{limit5_4}
\end{equation}
interchanging $3$ and $4$ in \eqref{limit5_3}, subtracting, and using \eqref{limit5_4}, we get
\begin{equation}
[\sigma_{;12345}] = [\sigma_{;12435}];
\end{equation}
thus \eqref{limit5_3} becomes
\begin{eqnarray}
&&[\sigma_{;12345}] = -\tfrac{1}{4}( (1,3,2,4;5)+(1,3,2,5;4) +(1,4,2,3;5) 
\nonumber \\
&&\;\;\;\;\;+(1,4,2,5;3)+ (1,5,2,3;4) + (1,5,2,4;3)  ) .
\end{eqnarray}
or
\begin{equation}
[\sigma_{;12345}] = -\tfrac{1}{4}  \mathcal{P}_{3,4,5} (1,3,2,4;5).
\end{equation}

Fo the sixth order we have, by \eqref{fund1_gen},
\begin{eqnarray}
&&[\sigma_{;162345}]+[\sigma_{;152346}]+[\sigma_{;142356}] 
\nonumber \\
&& \;\;\;\;\;+[\sigma_{;132456}]+[\sigma_{;123456}]+ [H_{123456}] = 0, 
\label{limit6_1}
\end{eqnarray}
where
\begin{eqnarray}
&& [H_{123456}] 
\nonumber \\
&&\;\;\;\;\;= [\tensor{\sigma}{_{;1\lambda 23}}][\tensor{\sigma}{^{;\lambda} _{456}}] 
+ [\tensor{\sigma}{_{;1\lambda 24}}][\tensor{\sigma}{^{;\lambda} _{356}}] 
+[\tensor{\sigma}{_{;1\lambda 25}}][\tensor{\sigma}{^{;\lambda} _{346}}]
+[\tensor{\sigma}{_{;1\lambda 26}}][\tensor{\sigma}{^{;\lambda} _{345}}] 
\nonumber \\
&& \;\;\;\;\; + [\tensor{\sigma}{_{;1\lambda 34}}][\tensor{\sigma}{^{;\lambda} _{256}}]
+[\tensor{\sigma}{_{;1\lambda 35}}][\tensor{\sigma}{^{;\lambda} _{246}}]
+[\tensor{\sigma}{_{;1\lambda 36}}][\tensor{\sigma}{^{;\lambda} _{245}}] 
\nonumber \\ 
&& \;\;\;\;\; + [\tensor{\sigma}{_{;1\lambda 45}}][\tensor{\sigma}{^{;\lambda} _{236}}]
+[\tensor{\sigma}{_{;1\lambda 46}}][\tensor{\sigma}{^{;\lambda} _{235}}] 
\nonumber \\
&& \;\;\;\;\; +[\tensor{\sigma}{_{;1\lambda 56}}][\tensor{\sigma}{^{;\lambda} _{234}}] 
\label{H_def}
\end{eqnarray}
is a tensor whose value is known by \eqref{limit4_res}; from \eqref{fund2_gen} we have
\begin{eqnarray}
&&[\sigma_{;123456}]-[\sigma_{;132456}] 
\nonumber \\
&& \;\;\;\;\; = -(1,4,2,3;5,6)-(1,5,2,3;4,6) 
\nonumber \\
&& \;\;\;\;\; -(1,6,2,3;4,5)-(1,\lambda ,2,3) [\tensor{\sigma}{^{;\lambda} _{4,5,6}} ]. 
\end{eqnarray}
By means of this equation we convert \eqref{limit6_1} into
\begin{eqnarray}
&&[\sigma_{;126345}]+[\sigma_{;125346}]+[\sigma_{;124356}]+2[\sigma_{;123456}]+[H_{123456}] 
\nonumber \\
&& \;\;\;\;\; +(1,4,2,3;5,6)+(1,5,2,3;4,6)+(1,6,2,3;4,5) + (1,\lambda ,2,3)
\tensor{\sigma}{^{;\lambda} _{456}} 
\nonumber \\
&& \;\;\;\;\; +(1,3,2,4;5,6)+(1,5,2,4;3,6)+(1,6,2,4;3,5) + (1,\lambda ,2,4)
\tensor{\sigma}{^{;\lambda} _{356}} 
\nonumber \\
&& \;\;\;\;\; +(1,3,2,5;4,6)+(1,4,2,5;3,6)+(1,6,2,5;3,4) + (1,\lambda ,2,5)
\tensor{\sigma}{^{;\lambda} _{346}} 
\nonumber \\
&& \;\;\;\;\; +(1,3,2,6;4,5)+(1,4,2,6:3,5)+(1,5,2,3;6,4) + (1,\lambda ,2,6)
\tensor{\sigma}{^{;\lambda} _{345}} 
\nonumber \\
&& \;\;\;\;\;\;=0.
\label{limit6_2}
\end{eqnarray}
If we employ the symbol $I$ to denote the operation of interchanging numerals, so that 
$I(3,4)$, for example, denotes an interchange of $3$ and
$4$, the previous equation may be written
\begin{eqnarray}
&&\lbrace 2+I(3,4)+I(4,5)I(3,4)+I(5,6)I(4,5)I(3,4) \rbrace [\sigma_{;123456}] 
\nonumber \\
&& \;\;\;\;\;+\lbrace 1+I(3,4)+I(4,5)I(3,4)+I(5,6)I(4,5)I(3,4) \rbrace [L_{123456}] 
\nonumber \\
&& \;\;\;\;\;+ [H_{123456}] = 0, \label{limit6_2bis}
\end{eqnarray}
where $[L_{123456}]$ denotes the second line of \eqref{limit6_2}.

Operating with$\lbrace 1- I(5,6)\rbrace$ on \eqref{limit6_2}, and using \eqref{limit4_sym}, we get
\begin{eqnarray}
&&\lbrace 2+I(3,4)\rbrace \lbrace 1-I(5,6)\rbrace [\sigma_{;123456}] 
\nonumber \\
&& \;\;\;\;\; +\lbrace 1+I(3,4)\rbrace \lbrace 1-I(5,6)\rbrace (1,3,2,4;5,6) = 0.  
\label{limit6_3}
\end{eqnarray}
Since $$I(3,4)I(3,4)=1,$$ we have
\begin{eqnarray}
&& \lbrace 2-I(3,4)\rbrace \lbrace 2+I(3,4)\rbrace = 3 ,  
\nonumber \\
&& \lbrace 2-I(3,4)\rbrace \lbrace 1+I(3,4)\rbrace = 1 + I(3,4) ,
\end{eqnarray}
and therefore operation on \eqref{limit6_3} with $\lbrace 2-I(3,4)\rbrace$ gives 
\begin{equation}
\lbrace 1-I(5,6)\rbrace [\sigma_{;123456}]=-\tfrac{1}{3}\lbrace 1
+I(3,4)\rbrace\lbrace 1-I(5,6)\rbrace (1,3,2,4;5,6) . 
\label{limit6_4}
\end{equation}
Operation on \eqref{limit6_2} with $\lbrace 1-I(4,5)\rbrace$ gives
\begin{eqnarray}
&&\lbrace 1-I(4,5)\rbrace [\sigma_{;126345}]+2\lbrace 1-I(4,5)
\rbrace[\sigma_{;123456}]+\lbrace 1-I(4,5)\rbrace [H_{123456}] 
\nonumber \\
&&\;\;\;\;\;+\lbrace 1+I(3,6)\rbrace\lbrace 1-I(4,5)\rbrace(1,3,2,6;4,5)
\nonumber \\
&&\;\;\;\;\;+\lbrace 1-I(4,5)\rbrace (1,\lambda ,2,3) [\tensor{\sigma}{^{;\lambda} _{456}}]=0, 
\label{limit6_5}
\end{eqnarray}
in which the first term may be evaluated by applying to \eqref{limit6_4} the substitution
\begin{eqnarray}
\;\;\;\;\; 1 \;\;2 \;\;3 \;\;4 \;\;5 \;\; 6 
\nonumber \\
\;\;\;\;\; 1 \;\;2 \;\;6 \;\;3 \;\;4 \;\; 5 
\nonumber
\end{eqnarray}
while the term in $[H]$ is, by \eqref{H_def}
\begin{equation}
\lbrace 1-I(4,5)\rbrace [\tensor{\sigma}{_{;1\lambda 23}}][\tensor{\sigma}{^{;\lambda} _{456}}].
\end{equation}
Thus \eqref{limit6_5} may be written
\begin{eqnarray}
&&\lbrace 1-I(4,5)\rbrace [\tensor{\sigma}{_{;123456}}] = -\tfrac{1}{3}\lbrace 1+I(3,6)\rbrace 
\lbrace 1-I(4,5)\rbrace (1,3,2,6;4,5) 
\nonumber \\
&& \;\;\;\;\; -\tfrac{1}{2}\lbrace 1-I(4,5)\rbrace ((1,\lambda ,2,3)
+ [\tensor{\sigma}{_{;1\lambda 23}}][\tensor{\sigma}{^{;\lambda} _{456}}]. 
\label{limit6_6}
\end{eqnarray}
Operating on \eqref{limit6_2} with $\lbrace 1-I(3,4)\rbrace$ gives
\begin{eqnarray}
&&\lbrace 1+I(5,6)\rbrace\lbrace 1-I(3,4)\rbrace [\tensor{\sigma}{_{;125346}}]
+\lbrace 1-I(3,4)\rbrace [\tensor{\sigma}{_{;123456}}] 
\nonumber \\
&&\;\;\;\;\; +\lbrace 1-I(3,4)\rbrace [H_{123456}] 
\nonumber \\
&&\;\;\;\;\; +\lbrace 1+I(5,6)\rbrace \lbrace 1-I(3,4)\rbrace (1,5,2,6;3,4) 
\nonumber \\
&&\;\;\;\;\; +\lbrace 1+I(5,6)\rbrace \lbrace 1-I(3,4)\rbrace (1,\lambda , 2,5)
[\tensor{\sigma}{^{;\lambda} _{346}}] = 0. 
\label{limit6_7}
\end{eqnarray}
The first term may be evaluated by applying to \eqref{limit6_6} the substitution
\begin{eqnarray}
\;\;\;\;\; 1 \;\;2 \;\;3 \;\;4 \;\;5 \;\; 6 
\nonumber \\
\;\;\;\;\; 1 \;\;2 \;\;5 \;\;3 \;\;4 \;\; 6 
\nonumber
\end{eqnarray}
and it becomes
\begin{eqnarray}
&& -\tfrac{1}{3}\lbrace 1+I(5,6)\rbrace \lbrace 1+I(5,6)\rbrace
\lbrace 1-I(3,4)\rbrace (1,5,2,6;3,4) 
\nonumber \\
&&\;\;\;\;\; -\tfrac{1}{2}\lbrace 1+I(5,6)\rbrace\lbrace 1-I(3,4)\rbrace 
((1,\lambda ,2,5) +[\tensor{\sigma}{_{;1\lambda 25}}])[\tensor{\sigma}{^{;\lambda}_{346}}],
\end{eqnarray}
in which the first term can be simplified, since
\begin{equation}
\lbrace 1+I(5,6)\rbrace \lbrace 1+I(5,6)\rbrace = 2 \lbrace 1+I(5,6)\rbrace .
\end{equation}
The $[H]$ term in \eqref{limit6_7} is, by \eqref{H_def}
\begin{eqnarray}
\lbrace 1+I(5,6)\rbrace \lbrace 1-I(3,4)\rbrace 
[\tensor{\sigma}{_{;1\lambda 25}}][\tensor{\sigma}{^{;\lambda}_{346}}].
\end{eqnarray}
Thus \eqref{limit6_7} may be written
\begin{eqnarray}
&&\lbrace 1-I(3,4)\rbrace  [\tensor{\sigma}{_{;123456}}] 
= -\tfrac{1}{3}\lbrace 1+I(5,6)\rbrace 
\lbrace 1-I(3,4)\rbrace (1,5,2,6;3,4) 
\nonumber \\
&& \;\;\;\;\; -\tfrac{1}{2}\lbrace 1+I(5,6)\rbrace \lbrace 1-I(3,4)\rbrace 
((1,\lambda ,2,5) +[\tensor{\sigma}{_{;1\lambda 25}}])[\tensor{\sigma}{^{;\lambda}_{346}}]. 
\label{limit6_8}
\end{eqnarray}
We have now to apply \eqref{limit6_4}, \eqref{limit6_6}, \eqref{limit6_8} to the solution 
of \eqref{limit6_2} or \eqref{limit6_2bis}. The first term in \eqref{limit6_2bis} is
\begin{eqnarray}
&& \lbrace 2 + (l+I(4, 5)+I(5, 6)I(4, 5)) I(3,4)\rbrace [\sigma_{;123456}] 
\nonumber \\
&&\;\;\;\;\; = \lbrace 3+ (1+I(5,6))I(4,5) \rbrace  [\sigma_{;123456}] 
\nonumber \\
&& \;\;\;\;\; - \lbrace 1+I(4,5)+I(5,6)I(4,5) \rbrace 
\lbrace 1-I(3,4) \rbrace [\sigma_{;123456}];
\end{eqnarray}
but 
\begin{eqnarray}
&& \lbrace 3+ (1+I(5,6))I(4,5) \rbrace  [\sigma_{;123456}] 
\nonumber \\
&&  =\lbrace 4+I(5,6) \rbrace [\sigma_{;123456}] - \lbrace 1+I(5,6) \rbrace 
\lbrace 1- I(4,5) \rbrace  [\sigma_{;123456}]
\end{eqnarray}
and $$\lbrace 4+I(5,6) \rbrace [\sigma_{;123456}] = 5 [\sigma_{;123456}] 
- \lbrace 1-I(5,6) \rbrace [\sigma_{;123456}].$$ After adding these three equations, and using 
\eqref{limit6_4}, \eqref{limit6_6}, \eqref{limit6_8}, 
we find that the first term of \eqref{limit6_2bis} is 
\begin{eqnarray}
&& 5[\sigma_{;123456}] +\tfrac{1}{3}\lbrace 1+I(3,4) \rbrace 
\lbrace 1-I(5,6) \rbrace (1,3,2,4;5,6) 
\nonumber \\
&& \;\;\;\;\; +\tfrac{1}{3}\lbrace 1+I(5,6) \rbrace \lbrace 1+I(3,6) \rbrace
\lbrace 1-I(4,6) \rbrace (1,3,2,6;4,5) 
\nonumber \\
&& \;\;\;\;\; +\tfrac{1}{2}\lbrace 1+I(5,6) \rbrace\lbrace 1-I(4,5) \rbrace 
((1,\lambda ,2,3)+\tensor{\sigma}{_{;1\lambda 23}})\tensor{\sigma}{^{;\lambda} _{456}} 
\nonumber \\
&& \;\;\;\;\; +\tfrac{1}{3} \lbrace 1+I(4,5) +I(5,6)I(4,5) \rbrace 
\lbrace 1+I(5,6) \rbrace \lbrace 1-I(3,4) \rbrace (1,5,2,3;6,4) 
\nonumber \\
&& \;\;\;\;\; +\tfrac{1}{2} \lbrace 1+I(4,5) +I(5,6)I(4,5) \rbrace 
\lbrace 1+I(5,6) \rbrace \lbrace 1-I(3,4) \rbrace \bullet 
\nonumber \\
&& \;\;\;\;\;\;\;\;\;\; \bullet ((1,\lambda ,2,5)+\tensor{\sigma}{_{;1\lambda 25}})
\tensor{\sigma}{^{;\lambda} _{346}}.
\end{eqnarray}
Substituting this expression for the first term in \eqref{limit6_2bis}, we get
\begin{eqnarray}
&& 5[\sigma_{;123456}] 
\nonumber \\
&& \;\;\;\;\;= - (1,4,2,3;5,6) -(1,5,2,3;4,6) - (1,6,2,3;4,5) 
- (1,\lambda , 2,3)[\tensor{\sigma}{^{;\lambda} _{456}}] 
\nonumber \\
&& \;\;\;\;\; - (1,3,2,4;5,6) -(1,5,2,4;3,6) - (1,6,2,4;3,5) - (1,\lambda , 2,4)
[\tensor{\sigma}{^{;\lambda} _{356}}] 
\nonumber \\
&& \;\;\;\;\; - (1,3,2,5;4,6) -(1,4,2,5;3,6) - (1,6,2,5;3,4) - (1,\lambda , 2,5)
[\tensor{\sigma}{^{;\lambda} _{346}}] 
\nonumber \\
&& \;\;\;\;\; - (1,3,2,6;4,5) -(1,4,2,6;3,5) - (1,5,2,6;3,4) - (1,\lambda , 2,6)
[\tensor{\sigma}{^{;\lambda} _{345}}] 
\nonumber \\
&& \;\;\;\;\; -\tfrac{1}{3} ((1,3,2,4;5,6)-(1,3,2,4;6,5)+(1,3,2,6;4,5)-(1,3,2,6;5,4) 
\nonumber \\
&& \;\;\;\;\;\;\;\;\;\;\; +(1,4,2,3;5,6)-(1,4,2,3;6,5)+(1,6,2,3;4,5)-(1,6,2,3;5,4) 
\nonumber \\
&& \;\;\;\;\;\;\;\;\;\;\; +(1,3,2,5;4,6)-(1,3,2,5;6,4)+(1,5,2,6;3,4)-(1,5,2,6;4,3) 
\nonumber \\
&& \;\;\;\;\;\;\;\;\;\;\; +(1,5,2,3;4,6)-(1,5,2,3;6,4)+(1,6,2,5;3,4)-(1,6,2,5;4,3) 
\nonumber \\
&& \;\;\;\;\;\;\;\;\;\;\; +(1,4,2,6;3,5)-(1,4,2,6;5,3)+(1,4,2,5;3,6)-(1,4,2,5;6,3) 
\nonumber \\
&& \;\;\;\;\;\;\;\;\;\;\; +(1,6,2,4;3,5)-(1,6,2,4;5,3)+(1,5,2,4;3,6)-(1,5,2,4;6,3)) 
\nonumber \\
&& \;\;\;\;\; -\tfrac{1}{2} ( (1,\lambda ,2,3)+[\tensor{\sigma}{_{;1\lambda 23}}])
([\tensor{\sigma}{^{;\lambda} _{456}}]-[\tensor{\sigma}{^{;\lambda} _{546}}]
+[\tensor{\sigma}{^{;\lambda} _{465}}]-[\tensor{\sigma}{^{;\lambda} _{645}}] ) 
\nonumber \\
&& \;\;\;\;\; -\tfrac{1}{2} ( (1,\lambda ,2,4)+[\tensor{\sigma}{_{;1\lambda 24}}])
([\tensor{\sigma}{^{;\lambda} _{356}}]-[\tensor{\sigma}{^{;\lambda} _{536}}]
+[\tensor{\sigma}{^{;\lambda} _{365}}]-[\tensor{\sigma}{^{;\lambda} _{356}}] ) 
\nonumber \\
&& \;\;\;\;\; -\tfrac{1}{2} ( (1,\lambda ,2,5)+[\tensor{\sigma}{_{;1\lambda 25}}])
([\tensor{\sigma}{^{;\lambda} _{346}}]-[\tensor{\sigma}{^{;\lambda} _{436}}]
+[\tensor{\sigma}{^{;\lambda} _{364}}]-[\tensor{\sigma}{^{;\lambda} _{634}}] ) 
\nonumber \\
&& \;\;\;\;\; -\tfrac{1}{2} ( (1,\lambda ,2,6)+[\tensor{\sigma}{_{;1\lambda 26}}])
([\tensor{\sigma}{^{;\lambda} _{345}}]-[\tensor{\sigma}{^{;\lambda} _{435}}]
+[\tensor{\sigma}{^{;\lambda} _{354}}]-[\tensor{\sigma}{^{;\lambda} _{534}}] ) 
\nonumber \\
&& \;\;\;\;\; -[H_{123456}]. 
\label{limit6_10}
\end{eqnarray}
The last term in the first line, the terms in the eleventh line, and the first term in 
$-[H_{123456}]$ (see eq. \eqref{H_def}), together make up 
(by use \eqref{limit4_sym} and \eqref{limit4_res})
\begin{eqnarray}
&&\lbrace (1,\lambda ,2,3) + [\tensor{\sigma}{_{;1 \lambda 23}}] \rbrace 
\lbrace -2[\tensor{\sigma}{^{;\lambda} _{456}}]+\tfrac{1}{2}([\tensor{\sigma}{^{;\lambda} _{546}}]
+[\tensor{\sigma}{^{;\lambda} _{645}}]) 
\nonumber \\
&&\;\; =\lbrace -(2,3,\lambda ,1)-\tfrac{1}{3}((1,2,\lambda ,3)+(1,3,\lambda ,2))\rbrace \bullet 
\nonumber \\
&& \;\;\;\; \bullet \; \lbrace \tfrac{2}{3}g^{\lambda\mu}((\mu ,5,4,6)+(\mu ,6,4,5)) 
-\tfrac{1}{6}g^{\lambda\mu}((\mu ,6,5,4)+(\mu ,5,6,4)) \rbrace 
\nonumber \\
&&\;\; =-\tfrac{5}{6} g^{\lambda\mu} \lbrace (2,3,\lambda ,1)+\tfrac{1}{3} (1,2,\lambda ,3)
+\tfrac{1}{3}(1,3,\lambda ,2)\rbrace\bullet 
\nonumber \\
&& \;\;\;\; \bullet \; \lbrace (\mu ,5,4,6)+(\mu ,6,4,5)\rbrace 
\nonumber \\
&&\;\;=-\tfrac{5}{6} g^{\lambda\mu} \lbrace (2,3,\lambda ,1)+\tfrac{1}{3} 
(1,2,\lambda ,3)+\tfrac{1}{3}(1,3,\lambda ,2)\rbrace \bullet 
\nonumber \\
&& \;\;\;\; \bullet \;\lbrace (4,5,\mu ,6)+(4,6,\mu ,5)\rbrace . 
\label{limit6_9}
\end{eqnarray}
But $$ (2,3,\lambda ,1) + (2, \lambda ,1,3) + (2, 1,3,\lambda) = 0 $$ and therefore 
\begin{equation}
(2,3,\lambda ,1)= - (2, \lambda ,1,3) - (2, 1,3,\lambda),
\end{equation}
so that \eqref{limit6_9} is equal to
\begin{equation}
-\tfrac{5}{9} g^{\lambda\mu} \lbrace 2(1,3,\lambda ,2)-(1,2,\lambda ,3) \rbrace 
\lbrace (4,5,\mu ,6) + (4,6, \mu , 5) \rbrace .
\end{equation} 
The first of the remaining terms in $-[H_{123456}]$ which are not used in forming 
the previous expression and the three expressions similar to it, is (see \eqref{H_def}) equal to 
\begin{eqnarray}
-\tfrac{1}{9} g^{\lambda\mu} \lbrace (1,3,\lambda ,4) + (1,4,\lambda ,3) \rbrace 
\lbrace (2,3,\mu ,6)+(2,6,\mu ,3) \rbrace . 
\end{eqnarray} 
Thus we see that \eqref{limit6_10} may be written 
\begin{eqnarray}
&&[\sigma_{;123456}]  
\nonumber \\
&&\;\;\;\;\;\;\;= -\tfrac{4}{15} \lbrace (1,3,2,4;5,6) + (1,3,2,5;4,6) + (1,3,2,6;4,5) 
\nonumber \\
&&\;\;\;\;\;\;\;\;\;\; + (1,4,2,3;5,6) + (1,4,2,5;3,6) + (1,4,2,6;3,5) 
\nonumber \\
&&\;\;\;\;\;\;\;\;\;\; + (1,5,2,3;4,6) + (1,5,2,4;3,6) + (1,5,2,6;3,4) 
\nonumber \\
&&\;\;\;\;\;\;\;\;\;\; + (1,6,2,3;4,5) + (1,6,2,4;3,5) + (1,6,2,5;3,4) \rbrace 
\nonumber \\
&&\;\;\;\;\;\;\;\;\;\; -\tfrac{1}{15} \lbrace (1,3,2,4;6,5) + (1,3,2,5;6,4) + (1,3,2,6;5,4) 
\nonumber \\
&&\;\;\;\;\;\;\;\;\;\; + (1,4,2,3;6,5) + (1,4,2,5;6,3) + (1,4,2,6;5,3) 
\nonumber \\
&&\;\;\;\;\;\;\;\;\;\; + (1,5,2,3;6,4) + (1,5,2,4;6,3) + (1,5,2,6;4,3) 
\nonumber \\
&&\;\;\;\;\;\;\;\;\;\; + (1,6,2,3;5,4) + (1,6,2,4;5,3) + (1,6,2,5;4,3)  \rbrace 
\nonumber \\
&&\;\;\;\;\;\;\;\;\;\; -\tfrac{1}{9}g^{\lambda\mu} 
\sum_{i=1}^{4} A^{(i)}_{\lambda\mu \; 123456} 
\nonumber \\
&&\;\;\;\;\;\;\;\;\;\; -\tfrac{1}{45}g^{\lambda\mu}
\sum_{j=1}^{6} B^{(i)}_{\lambda\mu \; 123456},
\end{eqnarray}
where
\begin{eqnarray}
&& A^{(1)}_{\lambda\mu \; 123456}= ( 2(1,3,\lambda ,2) 
-(1,2,\lambda ,3))((4,5,\mu ,6)+(4,6,\mu ,5)) , 
\nonumber \\
&& A^{(2)}_{\lambda\mu \; 123456}= ( 2(1,4,\lambda ,2) 
-(1,2,\lambda ,4))((3,5,\mu ,6)+(3,6,\mu ,5)) ,
\nonumber\\
&& A^{(3)}_{\lambda\mu \; 123456}= ( 2(1,5,\lambda ,2) 
-(1,2,\lambda ,5))((3,4,\mu ,6)+(3,6,\mu ,4)) ,
\nonumber \\
&& A^{(4)}_{\lambda\mu \; 123456}= ( 2(1,6,\lambda ,2) 
-(1,2,\lambda ,6))((3,4,\mu ,5)+(3,5,\mu ,4)) ,
\nonumber \\
&& B^{(1)}_{\lambda\mu \; 123456}= ( (1,3,\lambda ,4) 
+(1,4,\lambda ,3))((2,5,\mu ,6)+(2,6,\mu ,5)), 
\nonumber\\
&& B^{(2)}_{\lambda\mu \; 123456}= ( (1,3,\lambda ,5) 
+(1,5,\lambda ,3))((2,4,\mu ,6)+(2,6,\mu ,4)) ,
\nonumber\\
&& B^{(3)}_{\lambda\mu \; 123456}= ( (1,3,\lambda ,6) 
+(1,6,\lambda ,3))((2,4,\mu ,5)+(2,5,\mu ,4)) ,
\nonumber\\
&& B^{(4)}_{\lambda\mu \; 123456}= ( (1,4,\lambda ,5) 
+(1,5,\lambda ,4))((2,3,\mu ,6)+(2,6,\mu ,3)),
\nonumber\\
&& B^{(5)}_{\lambda\mu \; 123456}= ( (1,4,\lambda ,6) 
+(1,6,\lambda ,4))((2,3,\mu ,5)+(2,5,\mu ,3)), 
\nonumber\\
&& B^{(6)}_{\lambda\mu \; 123456}= ( (1,5,\lambda ,6) 
+(1,6,\lambda ,5))((2,3,\mu ,4)+(2,4,\mu ,3)).
\nonumber
\end{eqnarray}

\section{Coefficient Bivectors: Coincidence Limits}
\setcounter{equation}{0}

We will derive the coincidence limits of the following coefficient bivectors: 
$[b_{1 \; \mu\nu'}]$, $[b_{1 \; \mu\nu' ;\rho}]$, $[b_{1 \; \mu\nu' ;\rho\omega}]$, 
$[b_{2 \; \mu\nu'}]$; these quantities, together with the fundamental formula for the 
derivatives of the coincidence limit of bitensors $$[T_{\alpha_1 \alpha_2 ... \beta' _1 
\beta' _2 ...}(x,x')]_{;\gamma} = [T_{\alpha_1 \alpha_2 ... \beta' _1 \beta' _2 ... 
{;\gamma'}}(x,x')] + [T_{\alpha_1 \alpha_2 ... \beta' _1 \beta' _2 ... {;\gamma}}(x,x')]$$ 
make it possible to evaluate explicitly the right-hand side in \eqref{div_part}.

\subsection{Coincidence Limit of $b_{1 \; \mu\nu'}$}

Consider the recurrence relation \eqref{rec_rel_bn} for the coefficient 
bivectors $\tensor{b}{_{n \; \mu \nu '}}$:
\begin{equation}
\sigma^{;\lambda}b_{n \; \mu\nu';\lambda} + n b_{n \; \mu\nu'} 
= \frac{1}{\sqrt{\Delta}}\left( \sqrt{\Delta}b_{n-1 \; \mu\nu'}\right)
\tensor{}{_{;\lambda}^\lambda} - R^{\lambda}_\mu b_{n-1 \; \lambda\nu'}.
\end{equation}
Since $b_{0 \; \mu\nu'} = g_{\mu\nu'}$, for $n=1$ the previous equation is
\begin{eqnarray}
&& \sigma^{;\lambda}b_{1 \; \mu\nu';\lambda} +  b_{1 \; \mu\nu'} 
= \frac{1}{\sqrt{\Delta}}\left( \sqrt{\Delta}  g_{\mu\nu'}\right)
\tensor{}{_{;\lambda}^\lambda} - R^{\lambda}_\mu g_{ \lambda\nu'} 
\nonumber \\
&&= \frac{1}{\sqrt{\Delta}}\left( \sqrt{\Delta}\tensor{}{_{;\lambda}^\lambda}
g_{\mu\nu'} + 2\sqrt{\Delta}{_{;\lambda}}  g_{\mu\nu' ;}{^\lambda} 
+ \sqrt{\Delta}  g_{\mu\nu'}\tensor{}{_{;\lambda}^\lambda}\right)
- R^{\lambda}_\mu g_{ \lambda\nu'}. 
\nonumber \\ 
\label{n=1,rec_rel}
\end{eqnarray}
Taking the coincidence limit, and using (see Ref. \cite{relativity1960general})
\begin{eqnarray}
\left[\sigma^{;\lambda}\right] &=& 0,  
\label{obvcoincfirst}\\
\left[\sqrt{\Delta} \right] &=& 1, \\
\left[\sqrt{\Delta}_{;\mu}\right] &=& 0, \\ 
\left[\tensor{\sqrt{\Delta}}{_{;\mu\nu}}\right] &=& \tfrac{1}{6} R_{\mu\nu}, \\
\left[g_{\mu\nu'} \right]&=&g_{\mu\nu}, \\
\left[g_{\mu\nu' ;\alpha} \right] &=& 0, \\
\left[g_{\mu\nu' ;\alpha\beta}\right]&=& -\tfrac{1}{2}R_{\mu\nu\alpha\beta}, 
\label{obvcoinclast}
\end{eqnarray}
one obtains
\begin{equation}
[b_{1 \; \mu\nu'}] = \tfrac{1}{6} R g_{\mu\nu} - R_{\mu\nu}.
\end{equation}
\subsection{Coincidence Limit of $b_{1 \; \mu\nu' ;\rho}$}
Taking the first derivative of \eqref{n=1,rec_rel}, one obtains
\begin{eqnarray}
&&\tensor{\sigma}{^\lambda _\rho} \tensor{b}{_{1 \; \mu\nu';\lambda}} 
+ \tensor{\sigma}{^\lambda} \tensor{b}{_{1 \; \mu\nu';\lambda\rho}}
+\tensor{b}{_{1 \; \mu\nu';\rho}} = 
\nonumber \\
&&= \left(\frac{1}{\sqrt{\Delta}}\right)_{;\rho}\left( \sqrt{\Delta}
\tensor{}{_{;\lambda}^\lambda}g_{\mu\nu'} + 2\sqrt{\Delta}_{;\lambda}  
\tensor{g}{_{\mu\nu' ;} ^\lambda} + \sqrt{\Delta}  
g_{\mu\nu'}\tensor{}{_{;\lambda}^\lambda}\right) 
\nonumber \\
&& \; \; \; \; \;+ \frac{1}{\sqrt{\Delta}}\left( \sqrt{\Delta}
\tensor{}{_{;\lambda}^\lambda}g_{\mu\nu'} + 2\sqrt{\Delta}{_{;\lambda}}  
\tensor{g}{_{\mu\nu' ;}^\lambda} + \sqrt{\Delta}  g_{\mu\nu'}
\tensor{}{_{;\lambda} ^\lambda}\right)_{;\rho} 
\nonumber \\
&& \; \; \; \; \; - R^{\lambda}_{\mu ; \rho} g_{ \lambda\nu'}
- R^{\lambda}_\mu g_{ \lambda\nu' ;\rho}.
\end{eqnarray}
Taking the coincidence limit, and using \eqref{obvcoincfirst}-\eqref{obvcoinclast} and 
\begin{eqnarray}
\left[\left(\frac{1}{\sqrt{\Delta}}\right)_{;\rho}\right] &=& 
- \left[ \left(\sqrt{\Delta}\right)^{-2} \right] [ \sqrt{\Delta}_{;\rho} ] = 0, \\
\left[ \tensor{\sigma}{_{;\mu\nu}} \right] &=& g_{\mu\nu}, 
\end{eqnarray}
one obtains 
\begin{eqnarray}
2 [ b_{1 \; \mu\nu' ;\rho}] = \left[\tensor{\sqrt{\Delta}}{_{;\lambda}^\lambda _\rho}\right] 
g_{\mu\nu} + \left[\tensor{g}{_\mu _{\nu';\lambda} ^\lambda _\rho}\right]
- R_{\mu\nu;\rho}. 
\label{n=1,coinclim1}
\end{eqnarray}
From 
$$
\left[\tensor{\sqrt{\Delta}}{_{;\alpha\beta\gamma}}\right]
= \tfrac{1}{12}\left(R_{\alpha\beta;\gamma}+R_{\alpha\gamma;\beta}
+R_{\beta\gamma;\alpha}\right),$$ $$\left[\tensor{g}{_\mu _{\nu';\alpha\beta\gamma}}\right]
= -\tfrac{1}{3}( R_{\mu\nu\alpha\beta;\gamma}+R_{\mu\nu\alpha\gamma;\beta}),
$$ 
one arrives at
\begin{eqnarray}
\left[\tensor{\sqrt{\Delta}}{_{;\lambda}^\lambda _\rho}\right] &=& 
\tfrac{1}{12}\left(R_{;\gamma}+2\tensor{R}{_{\lambda\rho;}^\lambda}\right), \\
\left[\tensor{g}{_\mu _{\nu';\lambda} ^\lambda _\rho}\right] &=& 
-\tfrac{1}{3}\tensor{R}{_{\mu\nu\alpha\gamma}^{;\alpha}}.
\end{eqnarray}
Then \eqref{n=1,coinclim1} reads
\begin{equation}
[b_{1 \; \mu\nu' ;\rho}] = \tfrac{1}{24}g_{\mu\nu}\left(R_{;\gamma}
+2\tensor{R}{_{\lambda\rho;}^\lambda}\right)-\tfrac{1}{6}
\tensor{R}{_{\mu\nu\alpha\gamma}^{;\alpha}}- \tfrac{1}{2}R_{\mu\nu;\rho}.
\end{equation}

\subsection{Coincidence Limit of $b_{1 \; \mu\nu' ;\rho\omega}$}

Upon differentiating again \eqref{n=1,rec_rel}, one obtains
\begin{eqnarray}
&&\tensor{\sigma}{_; ^\lambda _\rho _\omega} b_{1 \; \mu\nu'; \lambda} 
+ \tensor{\sigma}{_; ^\lambda _\rho } b_{1 \; \mu\nu'; \lambda\omega} 
+ \tensor{\sigma}{_; ^\lambda  _\omega} b_{1 \; \mu\nu'; \lambda\rho} 
+ \tensor{\sigma}{_; ^\lambda } b_{\ \; \mu\nu'; \lambda\rho\omega} 
+ b_{1 \; \mu\nu'; \rho\omega} 
\nonumber \\
&&= \left(\frac{1}{\sqrt{\Delta}}\right)_{;\rho\omega}\left( \sqrt{\Delta}
\tensor{}{_{;\lambda}^\lambda}g_{\mu\nu'} + 2\sqrt{\Delta}_{;\lambda}  
\tensor{g}{_{\mu\nu' ;} ^\lambda} + \sqrt{\Delta}  g_{\mu\nu'}
\tensor{}{_{;\lambda}^\lambda}\right) 
\nonumber \\
&& \; \; \; \; \; +\left(\frac{1}{\sqrt{\Delta}}\right)_{;\rho}
\left( \sqrt{\Delta}\tensor{}{_{;\lambda}^\lambda}g_{\mu\nu'} 
+ 2\sqrt{\Delta}_{;\lambda}  \tensor{g}{_{\mu\nu' ;} ^\lambda} 
+ \sqrt{\Delta}  g_{\mu\nu'}\tensor{}{_{;\lambda}^\lambda}\right)_{;\omega} 
\nonumber \\
&& \; \; \; \; \; + \left(\frac{1}{\sqrt{\Delta}}\right)_{;\omega}\left( 
\sqrt{\Delta}\tensor{}{_{;\lambda}^\lambda}g_{\mu\nu'} + 2\sqrt{\Delta}_{;\lambda}  
\tensor{g}{_{\mu\nu' ;} ^\lambda} + \sqrt{\Delta} g_{\mu\nu'}
\tensor{}{_{;\lambda}^\lambda}\right)_{;\rho}	
\nonumber \\
&& \; \; \; \; \; + \left(\frac{1}{\sqrt{\Delta}}\right)\left( 
\sqrt{\Delta}\tensor{}{_{;\lambda}^\lambda}g_{\mu\nu'} + 2\sqrt{\Delta}_{;\lambda}  
\tensor{g}{_{\mu\nu' ;} ^\lambda} + \sqrt{\Delta}  g_{\mu\nu'}
\tensor{}{_{;\lambda}^\lambda}\right)_{;\rho\omega} 
\nonumber \\
&& \; \; \; \; \; - \tensor{R}{_\mu ^\lambda _{;\rho\omega}}g_{\lambda\nu'} 
- \tensor{R}{_\mu ^\lambda _{;\rho}}g_{\lambda\nu';\omega} 
- \tensor{R}{_\mu ^\lambda _{;\omega}}g_{\lambda\nu';\rho}  
- \tensor{R}{_\mu ^\lambda }g_{\lambda\nu';\rho\omega}
\end{eqnarray}
Taking the coincidence limit, and using
\begin{eqnarray}
&&[ \tensor{\sigma}{_{;\alpha\beta\gamma}} ] = 0 \\
&&\left[ \Delta^{-1/2} _{;\alpha\beta} \right] 
= \left[ (-\Delta^{1/2} _{;\alpha\beta}/ \Delta )  
+ (2\Delta^{1/2} _{;\alpha}\Delta^{1/2} _{;\beta}/\Delta^{3/2}) \right] 
= -\tfrac{1}{6}R_{\alpha\beta}, 
\end{eqnarray}
one arrives at
\begin{eqnarray}
&& 2 [b_{1 \; \mu\nu'; \rho\omega}] +[b_{1 \; \mu\nu'; \omega\rho}] 
\nonumber \\
&&= -\tfrac{1}{36}R g_{\mu\nu} R_{\rho\omega} +\left[ 
\tensor{\Delta}{^{1/2} _{;\lambda}^\lambda _\rho _\omega} \right]g_{\mu\nu} 
-\tfrac{1}{12}R R_{\mu\nu\rho\omega} 
\nonumber \\
&& \;\;\;\;\; -\tfrac{1}{6} R^{\lambda}_\rho R_{\mu\nu\lambda\omega}
-\tfrac{1}{6} R^{\lambda}_\omega R_{\mu\nu\lambda\rho} 
+ [\tensor{g}{_{\mu\nu';} _\lambda^\lambda _\rho _\omega}]
\nonumber \\
&& \;\;\;\;\; - \tensor{R}{_\mu _\nu _{;\rho\omega}} +\tfrac{1}{2}  
\tensor{R}{_\mu ^\lambda}R_{\lambda\nu \rho\omega}. 
\label{n=1,firstder1}
\end{eqnarray}
Now we can exploit the close relation between the commutator of covariant derivatives 
acting on a (co-)vector and the Riemann tensor:
\begin{equation}
b_{1 \; \mu\nu'; \rho\omega}-b_{1 \; \mu\nu'; \omega\rho} 
= - \tensor{R}{_\mu ^\sigma _\rho _\omega}b_{1 \; \sigma\nu'};  
\end{equation}
hence, on taking the limit, we have
\begin{equation}
[b_{1 \; \mu\nu'; \rho\omega}]-[b_{1 \; \mu\nu'; \omega\rho}] 
= - \tensor{R}{_\mu ^\sigma _\rho _\omega}[b_{1 \; \sigma\nu'}].
\end{equation}
Inserting this equation in \eqref{n=1,firstder1}, one obtains
\begin{eqnarray}
&& 3[b_{1 \; \mu\nu'; \rho\omega}] 
= - \tensor{R}{_\mu ^\sigma _\rho _\omega}[b_{1 \; \sigma\nu'}] 
\nonumber \\
&& \; \; \; \; \; -\tfrac{1}{36}R g_{\mu\nu} R_{\rho\omega} +\left[ 
\tensor{\Delta}{^{1/2} _{;\lambda}^\lambda _\rho _\omega} \right]g_{\mu\nu} 
-\tfrac{1}{12}R R_{\mu\nu\rho\omega} 
\nonumber \\
&& \;\;\;\;\; -\tfrac{1}{6} R^{\lambda}_\rho R_{\mu\nu\lambda\omega}
-\tfrac{1}{6} R^{\lambda}_\omega R_{\mu\nu\lambda\rho} 
+ [\tensor{g}{_{\mu\nu';} _\lambda^\lambda _\rho _\omega}]
\nonumber  \\
&& \;\;\;\;\; - \tensor{R}{_\mu _\nu _{;\rho\omega}} +\tfrac{1}{2}  
\tensor{R}{_\mu ^\lambda}R_{\lambda\nu \rho\omega},
\end{eqnarray}
which can be fully expanded using
\begin{eqnarray}
&&S_{\lambda\mu\nu\rho} \equiv -\tfrac{1}{3}(R_{\lambda\nu\mu\rho}
+R_{\lambda\rho\mu\nu}), \\
&& [ \sqrt{\Delta}_{;\alpha\beta\gamma\delta}] = -\tfrac{1}{8}\lbrace 
[\tensor{\sigma}{^{;\rho} _{\rho\alpha\beta\gamma\delta}}] 
\nonumber \\
&& \;\;\;\;\; -\tfrac{1}{3}(R_{\alpha\rho}\tensor{R}{^\rho _{\beta\gamma\delta}} 
+R_{\beta\rho}\tensor{R}{^\rho _{\alpha\gamma\delta}}+R_{\gamma\rho}
\tensor{R}{^\rho _{\alpha\beta\delta}}+R_{\delta\rho}
\tensor{R}{^\rho _{\alpha\beta\gamma}}) 
\nonumber \\
&& \;\;\;\;\; +\tfrac{1}{3}(R_{\alpha\rho}\tensor{S}{^\rho _{\beta\gamma\delta}} 
+R_{\beta\rho}\tensor{S}{^\rho _{\alpha\gamma\delta}}+R_{\gamma\rho}
\tensor{S}{^\rho _{\alpha\beta\delta}}+R_{\delta\rho}
\tensor{S}{^\rho _{\alpha\beta\gamma}}) 
\nonumber \\
&& \;\;\;\;\; -\tfrac{2}{9}(R_{\alpha\beta}R_{\gamma\delta} 
+ R_{\alpha\gamma}R_{\beta\delta}+R_{\alpha\delta}R_{\beta\gamma}) \rbrace, 
\label{fourth_del} \\
&&[g_{\alpha\beta';\mu\nu\sigma\tau}] = -\tfrac{1}{4}(R_{\alpha\beta\mu\nu;\sigma\tau} 
+ R_{\alpha\beta\mu\sigma;\nu\tau}+ R_{\alpha\beta\mu\tau;\nu\sigma}) 
\nonumber \\
&& \; \; \; \; \; +\tfrac{1}{8}(\tensor{R}{_\alpha _\beta _\rho _\tau}
\tensor{S}{^\rho _\mu _\nu _\sigma}+\tensor{R}{_\alpha _\beta _\rho _\sigma}
\tensor{S}{^\rho _\mu _\nu _\tau}+\tensor{R}{_\alpha _\beta _\rho _\nu}
\tensor{S}{^\rho _\mu _\sigma _\tau}+ 
\nonumber \\
&& \; \; \; \; \; +\tensor{R}{_\alpha _\beta _\rho _\mu}
\tensor{S}{^\rho _\nu _\sigma _\tau}) 
\nonumber \\
&& \; \; \; \; \; -\tfrac{1}{8}(\tensor{R}{^\rho _\beta _\sigma _\tau}
\tensor{R}{_\rho _\alpha _\mu _\nu}+\tensor{R}{^\rho _\beta _\mu _\nu}
\tensor{R}{_\rho _\alpha _\sigma _\tau}+\tensor{R}{_\alpha _\beta _\rho _\tau}
\tensor{R}{^\rho  _\mu _\nu _\sigma} 
\nonumber \\
&& \; \; \; \; \; +\tensor{R}{_\alpha _\beta _ \rho _\sigma }\tensor{R}{^\rho _\mu _\nu _\tau} 
+ \tensor{R}{^\rho _\beta _\mu _\tau}\tensor{R}{_\rho _\alpha _\nu _\sigma}
+\tensor{R}{^\rho _\beta _\nu _\sigma}\tensor{R}{_\rho _\alpha _\mu _\tau} 
\nonumber \\
&& \; \; \; \; \; +\tensor{R}{^\rho _\beta _\nu _\tau}\tensor{R}{_\rho _\alpha _\mu _\sigma}
+\tensor{R}{^\rho _\beta _\mu _\sigma}\tensor{R}{_\rho _\alpha _\nu _\tau}
+\tensor{R}{_\alpha _\beta _\rho _\nu}\tensor{R}{^\rho _\mu _\sigma _\tau} 
\nonumber \\
&& \; \; \; \; \; +\tensor{R}{_\alpha _\beta _\mu _\rho}
\tensor{R}{^\rho _\nu _\sigma _\tau}), 
\label{fourth_transp_par}
\end{eqnarray}
\subsection{Coincidence Limit of $b_{2 \; \mu\nu'}$}
For $n=2$, eq. \eqref{rec_rel_bn} is:
\begin{eqnarray}
&&\sigma^{;\lambda}b_{2 \; \mu\nu';\lambda} + 2 b_{2 \; \mu\nu'} 
\nonumber \\
&&= \frac{1}{\sqrt{\Delta}}\left( \sqrt{\Delta}\tensor{}{_{;\lambda}^\lambda}b_{1 \;\mu\nu'} 
+ 2\sqrt{\Delta}{_{;\lambda}}  b_{1 \; \mu\nu' ;}{^\lambda} 
+ \sqrt{\Delta}  b_{1 \; \mu\nu'}\tensor{}{_{;\lambda}^\lambda}\right) 
- R^{\lambda}_\mu b_{1 \; \lambda\nu'}. 
\nonumber \\
\end{eqnarray}
Taking the coincidence limit, one obtains
\begin{eqnarray}
&&2 [b_{2 \; \mu\nu'}] = \tfrac{1}{6}R[b_{1 \;\mu\nu'}] 
+[b_{1 \; \mu\nu'}]\tensor{}{_{;\lambda}^\lambda}
- R^{\lambda}_\mu [b_{1 \; \lambda\nu'}] 
\nonumber\\
&&= (\tfrac{1}{36} R^2 g_{\mu\nu} - \tfrac{1}{6}R R_{\mu\nu}) 
+ \bigg\lbrace\tfrac{1}{3} g_{\mu\nu}\bigg(-\tfrac{1}{36}R^2 
+ \left[ \tensor{\sqrt{\Delta}}{^{1/2} _\alpha ^\alpha _\beta ^\beta}\right] \bigg) 
\nonumber \\
&& -\tfrac{1}{3} \Box R_{\mu\nu} + \tfrac{1}{3}\left[ 
\tensor{g}{_\mu _{\nu' ;\alpha} ^\alpha _\beta ^\beta}\right] \bigg\rbrace 
+( -\tfrac{1}{6}R R_{\mu\nu} + R^\lambda _\mu R_{\lambda\nu}). 
\label{coinc_lim_n=2_in}
\end{eqnarray}
By using \eqref{fourth_del}, \eqref{fourth_transp_par}, one arrives at
\begin{eqnarray}
\left[ \tensor{\sqrt{\Delta}}{^{1/2} _; _\alpha ^\alpha _\beta ^\beta}\right] 
&=& \tfrac{1}{5} \Box R + \tfrac{1}{36} R^2 -\tfrac{1}{30}R_{\alpha\beta}R^{\alpha\beta} 
+ \tfrac{1}{30}R_{\alpha\beta\gamma\delta} R^{\alpha\beta\gamma\delta}, \\
\left[ \tensor{g}{_\mu _{\nu' ;\alpha} ^\alpha _\beta ^\beta}\right] 
&=& -\tfrac{1}{2} \tensor{R}{^{\alpha\beta\gamma} _\mu}R_{\alpha\beta\gamma\nu}.
\end{eqnarray}
Hence \eqref{coinc_lim_n=2_in} reads
\begin{eqnarray}
&& 2 [b_{2 \; \mu\nu'}] = R_\mu ^\lambda R_{\lambda\nu} - \tfrac{1}{3}RR_{\mu\nu}
-\tfrac{1}{3} \Box R_{\mu\nu} -\tfrac{1}{6} 
\tensor{R}{^{\alpha\beta\gamma} _\mu}R_{\alpha\beta\gamma\nu} 
\nonumber \\
&& + (\tfrac{1}{36}R^2 +\tfrac{1}{15}\Box R -\tfrac{1}{90}R_{\alpha\beta}R^{\alpha\beta} 
+ \tfrac{1}{90}\tensor{R}{^{\alpha\beta\gamma\delta}}R_{\alpha\beta\gamma\delta})g_{\mu\nu}.
\end{eqnarray}

\end{appendix}

\end{document}